\pdfoutput=1

\documentclass[aps,prfluids,reprint,unsortedaddress,onecolumn]{revtex4-2}
\usepackage[export]{adjustbox}
\usepackage{geometry}
\usepackage{amsmath}
\usepackage{float}
\usepackage{subcaption}
\usepackage{xfrac}
\usepackage{lmodern}
\usepackage[hidelinks]{hyperref}
\usepackage{cleveref}
\usepackage{indentfirst}
\usepackage{graphicx}

\begin{document}

\title{From unsteady to quasi-steady dynamics in the streamwise-oscillating cylinder wake}

\author{Maysam Shamai}
\email[]{mshamai@caltech.edu}
\affiliation{Graduate Aerospace Laboratories, California Institute of Technology, Pasadena, CA 91125, USA}

\author{Scott T.M. Dawson}
\affiliation{Department of Mechanical, Materials, and Aerospace Engineering, Illinois Institute of Technology, Chicago, IL 60616}

\author{Igor Mezi\'{c}}
\affiliation{Department of Mechanical Engineering, University of California Santa Barbara, Santa Barbara, CA 93106, USA}

\author{Beverley J. McKeon}
\affiliation{Graduate Aerospace Laboratories, California Institute of Technology, Pasadena, CA 91125, USA}

\date{\today}

\begin{abstract}
	The flow around a cylinder oscillating in the streamwise direction with a frequency, $f_f$, much lower than the shedding frequency, $f_s$, has been relatively less studied than the case when these frequencies have the same order of magnitude, or the transverse oscillation configuration. In this study, Particle Image Velocimetry and Koopman Mode Decomposition are used to investigate the streamwise-oscillating cylinder wake for forcing frequencies $f_f/f_s \sim 0.04-0.2$ and mean Reynolds number, $Re_0 = 900$. The amplitude of oscillation is such that the instantaneous Reynolds number remains above the critical value for vortex shedding at all times.  Characterization of the wake reveals a range of phenomena associated with the interaction of the two frequencies, including modulation of both the amplitude and frequency of the wake structure by the forcing. Koopman analysis reveals a frequency spreading of Koopman modes. A scaling parameter and associated transformation are developed to relate the unsteady, or forced, dynamics of a system to that of a quasi-steady, or unforced, system. For the streamwise-oscillating cylinder, it is shown that this transformation leads to a Koopman Mode Decomposition similar to that of the unforced system.
\end{abstract}

\maketitle

\section{Introduction}
Fluid-structure interaction serves as the backdrop of many important problems involving flow physics. Even the presence of a simple structure in the flow leads to additional boundary conditions that can significantly alter the system's dynamics. The complexity of the system is further increased when the structure takes on motion, whether predefined or in the form of a response to the flow. Although a moving structure introduces an added level of complexity, often many of the flow structures encountered may also be seen for the fixed case. The difficulty, however, lies in determining which characteristics of the flow carry over and if so, are the underlying physics governing them related. In this work, the streamwise-oscillating cylinder is used to study this question and develop a framework to relate the moving and fixed fluid-structure problems. Specifically, theory, experiment, and reduced-order modeling are used to address various aspects of the problem.

An expansive body of work considers the flow around oscillating cylinders; examples include Williamson \& Roshko (1988) and Williamson \& Govardhan (2004). Examining the literature, however, reveals that only a small subset of studies consider streamwise oscillations. This is primarily due to significant interest in transverse oscillations and their role in vortex-induced vibration (VIV). Furthermore, of the body of work considering streamwise-oscillations, the majority of studies consider forcing frequencies, $f_f$, of the same order of magnitude (or higher) than the stationary shedding frequency, $f_0$. This regime is of interest due to the lock-on phenomenon, the synchronization between the shedding frequency, $f_s$, and forcing frequency, $f_f$, that occurs for certain frequency ratios when the streamwise forcing frequency is approximately equal to the stationary shedding frequency \cite{Barbi_1986}. Streamwise forcing not corresponding to lock-on can also lead to wake patterns that significantly deviate from the classic K\'{a}rm\'{a}n vortex street \cite{Barbi_1986},\cite{Griffin_1976},\cite{Leontini_2011},\cite{Leontini_2013}.

The phenomena that arise due to streamwise forcing are intricately related to vortex dynamics as well as the forcing frequency and amplitude. Leontini et al. (2011) studied the streamwise-oscillating cylinder using two-dimensional computations with mean Reynolds number, $Re_0 = 175$. They considered a forcing frequency equivalent to the stationary shedding frequency and deduced that the development of frequency modulation in the wake is governed by the forcing frequency while vortex interactions lead to the development of amplitude modulation. Leontini et al. (2013) parametrically studied the streamwise-oscillating cylinder for $75 \le Re_0 \le 250$ and forcing frequencies in the range $1\le f_f/f_0\le 2$.

Although the range of forcing frequencies corresponding to lock-on has received ample consideration, the effect of forcing frequencies one-to-two orders of magnitude less that the stationary shedding frequency has received much less attention, even though it corresponds to numerous fluid-structure systems, for example, vortex shedding from an airfoil undergoing dynamic stall. The current study examines forcing frequencies one and two orders-of-magnitude less than the stationary shedding frequency.

Modal decompositions have proven useful in both the extraction of dynamically significant flow structures as well as reduced order modeling \cite{Rowley_2017},\cite{Taira_2017}.  In this study Koopman analysis is used to extract dynamically significant flow structures. In brief, the Koopman operator can be used to study the dynamics of a finite dimensional nonlinear system within a linear, but infinite dimensional framework \cite{Rowley_2009}. The Koopman mode decomposition (KMD) specifically refers to the eigen-decomposition of the Koopman operator. The power of KMD lies in its ability to extract flow structures that are of spatio-temporal importance \cite{Rowley_2009}. Although the Koopman operator is infinite dimensional, Koopman modes and eigenvalues can be approximated from data using Dynamic Mode Decomposition \cite{Rowley_2009}. This method has been used extensively for a large range of systems, e.g. \cite{Tu_2011}.

Glaz et al. (2017) used Koopman analysis to analyze the temporal behavior of a streamwise-oscillating cylinder with mean Reynolds number, $Re_0 = 53$ (close to the critical value). They considered a forcing frequency two orders of magnitude less than the stationary shedding frequency for various oscillation amplitudes. Koopman analysis was performed in parallel to two-dimensional flow simulations and the results were in good agreement. Glaz et al. showed that the flow in this regime was marked by instances of strong vortex shedding which were then replaced by periods of suppressed shedding and weak oscillatory behavior, a phenomenon they refer to as ``quasi-periodic intermittency". Additionally, they observed that streamwise forcing led to the extraction of additional flow structures (modes) related to vortex shedding as well as a broadening of the associated spectral peak.

The remainder of this paper is organized as follows. In \cref{section:analysis}, a parameter is developed to assess the degree to which a forced periodic system can be approximated as exhibiting quasi-steady behavior. Following that, a scaling procedure is presented that can be used to transform the dynamics of an unsteady system to one exhibiting quasi-steady dynamics; this analysis is then applied to the streamwise-oscillating cylinder. In \cref{section:Methods}, the experimental approach, namely Particle Image Velocimetry, used to study the flow around the streamwise-oscillating cylinder is outlined. The framework used to compute Koopman modes and eigenvalues from experimental velocity fields is also discussed. The results of oscillating-cylinder experiments, the associated Koopman analysis and time scaling is presented in \cref{section:Results}. In \cref{section:Discussion} results are discussed in more detail and the implications of scaling on the data are highlighted. Finally, concluding remarks are given \cref{section:Conclusion}.

\section{Analysis}
\label{section:analysis}
The wide separation between the natural and forcing frequencies in the system under study motivates analysis of the interplay between unsteady and quasi-steady dynamics along a forcing trajectory. In order to identify the regions of parameter space where a quasi-steady analysis will be suitable, we develop first a quasi-steadiness parameter, then investigate a time scaling procedure that reduces the complexity of the forced flow field.

\subsection{Quasi-steadiness}\
Consider the effect of simple harmonic forcing with period $\tau_f$ on a general periodic system with a time varying characteristic frequency and period, $f(t)$ and $\tau(t)=\frac{1}{f(t)}$, respectively. In the case of the cylinder considered here, $f$ can be equated with the vortex shedding frequency. Performing a Taylor series expansion of the period around an arbitrary time $t = t_0$ for small $\Delta t = t-t_0$ gives
\begin{equation}
	\tau(t) = \tau(t_0) + \frac{d\tau}{dt} \bigg |_{t=t_0} (t-t_0) + O(t-t_0)^2
	\label{eqn:taylorExpansion}
\end{equation}
Writing $\dot{\tau} = \frac{d\tau}{dt}$, $\Delta \tau = \tau(t)-\tau(t_0)$, and rearranging yields
\begin{equation}
	\Delta t = \frac{\Delta \tau}{\dot{\tau} |_{t=t_0}}.
	\label{eqn:firstOrder}
\end{equation}
Conceptually, $\Delta t$ represents the amount of time at the current state in the forcing cycle required for a change in the characteristic period equal to $\Delta \tau$. Normalizing \cref{eqn:firstOrder} using the forcing period, $\tau_f$, and taking the absolute value gives a time-dependent measure of the closeness of the system to steady behavior in a linear sense:
\begin{equation}
	\Omega(t) \equiv \Bigg |\frac{\Delta t}{\tau_f} \Bigg |= \Bigg |\frac{\Delta \tau}{\mathrm{\hspace*{0.55cm}}\tau_f \dot{\tau} \big|_{t=t_0}}\Bigg |,
	\label{eqn:omega}
\end{equation}
where we will call $\Omega(t)$ the quasi-steadiness parameter. $\Omega(t)$ describes the number of forcing periods the system would need to spend at the current state (in the forcing cycle) in order to realize a change in the characteristic period corresponding to $\Delta \tau$.  For a system with extrema in period, $\tau_{max}$ and $\tau_{min}$, we may choose
\begin{equation}
	\Delta \tau = \tau_{max} - \tau_{min}.
	\label{eqn:deltaT_max}
\end{equation}
Then $\Omega(t) \gg 1$ corresponds to portions of the forcing trajectory where the temporal behavior of the system is evolving relatively slowly, the characteristic frequency remains almost constant, and hence the system can be considered to be quasi-steady. Many forcing periods are required to observe an appreciable change in the characteristic frequency. In contrast, a significant change in the characteristic frequency can be observed in a short amount of time when $\Omega(t) \ll 1$. That is, the system dynamics are evolving quickly and hence the system is unsteady. If the dependence of the characteristic frequency on a set of parameters, $x_1,...,x_n$, is known, such that $f = f(x_1,...,x_n)$, the time variation of the quasi-steadiness parameter can be computed a priori.

	The quasi-steadiness parameter also gives insight into the dominant frequencies displayed by the system along a forcing trajectory. Specifically, the dominant frequencies observed along a forcing trajectory correspond to peaks in $\Omega$. As will be seen, a cylinder subject to an oncoming freestream can be forced (translated) such that it displays quasi-steady behavior, specifically, vortex shedding at an approximately constant frequency, during different portions of a forcing cycle. But, the shedding frequency associated with the different quasi-steady sections of the forcing cycle, i.e. where $\Omega \gg 1$, may be different if the range of Reynolds numbers traversed is different.

\subsection{Time scaling for quasi-steady behavior}\label{section:timescaling}

We now investigate time scaling to enable the quasi-steady analysis of an unsteady system, with the objective of expressing a time-dependent characteristic frequency, $f(t)$, in terms of a constant frequency, $\tilde{f}(\tilde{t}) = \tilde{f}$. We use the periodicity of the forcing to require that
\begin{equation}
	\tilde{f} \mathrm{d}\tilde{t} = f(t)\mathrm{d}t,
	\label{eqn:differentialForm}
\end{equation}
Integration of \cref{eqn:differentialForm} gives
\begin{align}
	n                                     & = \int_{0}^{\tilde{t}}\tilde{f} \mathrm{d}t' = \int_{0}^{t} f(t') \mathrm{d} t' \\
	\iff \hspace{.8cm}\tilde{f} \tilde{t} & = \int_{0}^{t} f(t') \mathrm{d} t'                                              \\
	\iff \hspace{1cm}\tilde{t}            & = \frac{1}{\tilde{f}} \int_{0}^{t} f(t') \mathrm{d} t'
	\label{eqn:scaling}
\end{align}
\noindent
As an aside, it is observed that the form of \cref{eqn:scaling} resembles the expression used to compute the angle of a frequency modulated signal \cite{Haykin_2001}. Let $y(t)$ denote an observable of the system (e.g. velocity) and assume $y = y\big(f(t)\big)$; that is, the behavior of the observable is a function of the characteristic frequency. Rearranging \cref{eqn:differentialForm} gives
\begin{equation}
	\frac{\tilde{f}}{f(t)} =\frac{\mathrm{d} t}{\mathrm{d} \tilde{t}}
	\label{eqn:differentialScaling}
\end{equation}
Substitution of  \cref{eqn:differentialScaling} into the observable as follows gives:
\begin{equation}
	y\bigg(\frac{\mathrm{d} t}{\mathrm{d} \tilde{t}} f(t)\bigg) = y\big(\tilde{f}\big)
\end{equation}\\

Time scaling can be implemented by sampling data (either in experiments or post-processing) at a rate that is constant relative to scaled time, $\tilde{t}$. Doing so in post-processing requires temporal-interpolation of the time series. Varying the sampling frequency is akin to allowing the passage of time to contract or dilate so that the observed frequency remains constant, hence the term ``time scaling". The Quasi-steadiness, $\Omega$, can be used to predict quasi-steady behavior (i.e. roughly constant shedding frequency) during various portions of the streamwise forcing cycle. This allows for the use of time scaling to normalize the frequency exhibited by the system.  Similar to $\Omega$, if the dependence of $f$ on the forcing is known then $\tilde{t}$ can be computed a priori.

\section{Methods}
\label{section:Methods}

\subsection{Oscillating Cylinder Experiments}
\begin{figure}[H]
	\begin{center}
		\begin{subfigure}[b]{0.45\textwidth}
			\begin{center}
				\includegraphics[scale = 0.06]{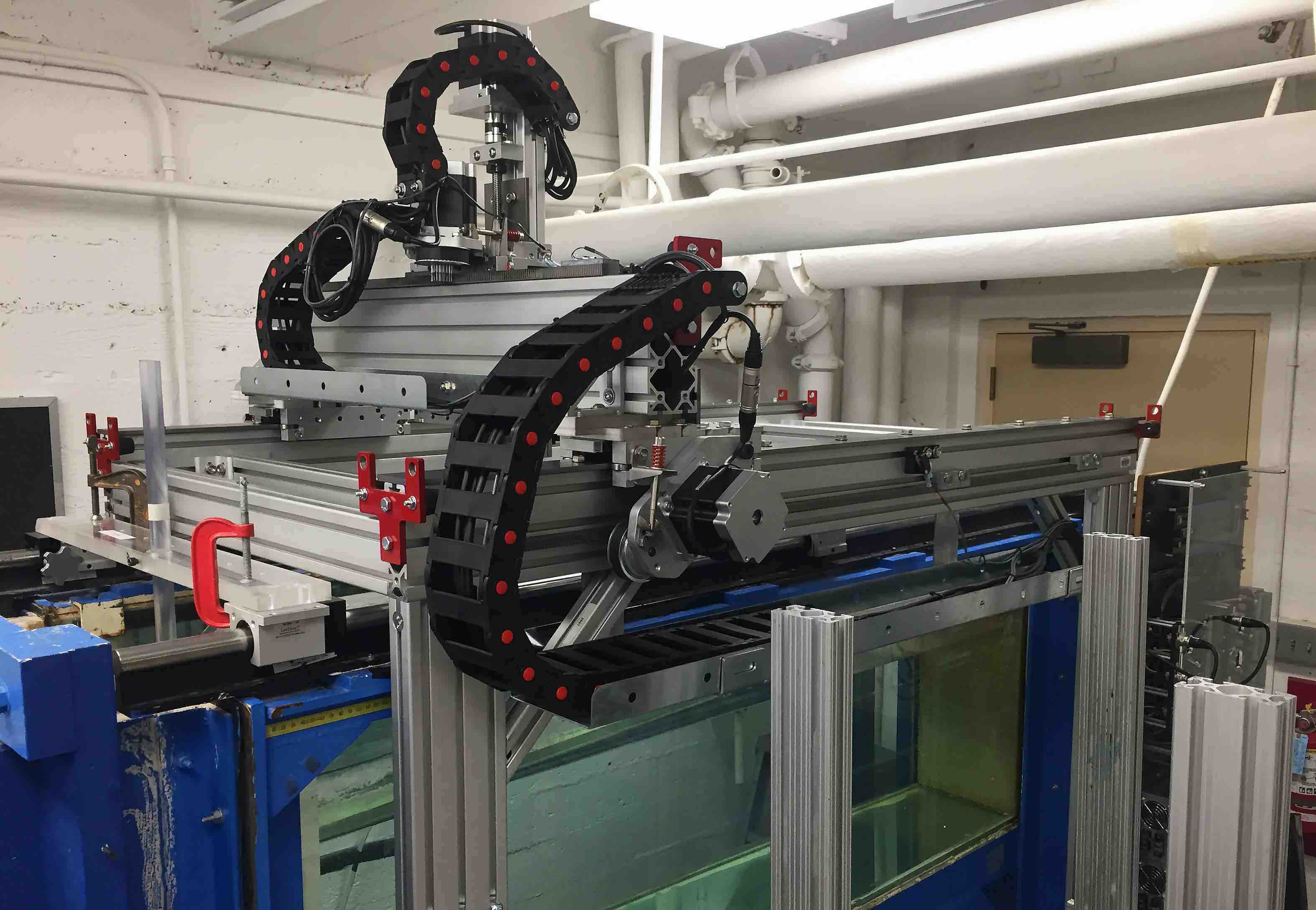}
				\caption{}
				\label{fig:CTS}
			\end{center}
		\end{subfigure}
		\begin{subfigure}[b]{0.34\textwidth}
			\begin{center}
				\includegraphics[trim =0px 0px 0px 0px, clip, scale = 0.62]{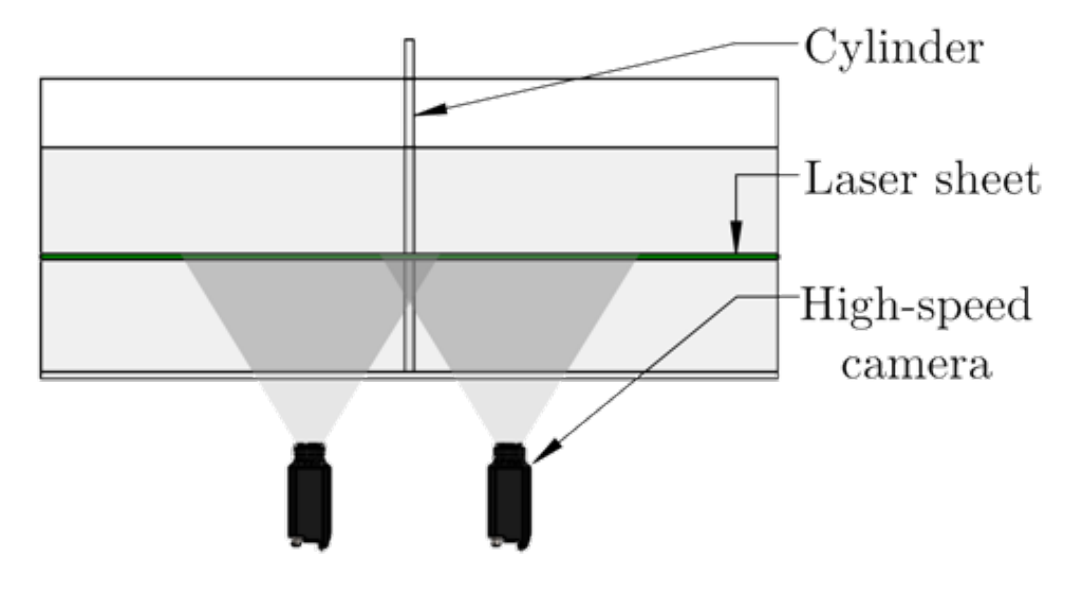}
				\caption{}
				\label{fig:Diagram}
			\end{center}
		\end{subfigure}
		\caption{(a) The captive trajectory system mounted on top of the NOAH water channel's test section. Flow is from left to right. (b) Simplified diagram of experimental setup.}
	\end{center}
\end{figure}

In order to study the flow around a streamwise-oscillating cylinder, experiments were carried out in the NOAH free-surface water channel at Caltech, a recirculating facility shown in \cref{fig:CTS}. A Captive Trajectory System (CTS) allows for translation of a model along all three coordinate axes as well as rotation about two of those axes. The CTS spans the length of the test section, approximately two meters in length, and can move with sub-millimeter accuracy reaching speeds up to 1 meter per second. In this study a prescribed trajectory is used to generate streamwise-oscillations.
\vspace{-0.5cm}

An acrylic cylinder with a diameter of 19mm and wetted length of approximately half a meter was mounted on the CTS such that its axis pointed downwards towards the bottom of the test section. In order to reduce end effects, the bottom of the cylinder was located approximately one millimeter from the bottom of the test section.

Two-dimensional particle image velocimetry (PIV) was the primary diagnostic used to acquire velocity fields. A Photonics DM20-527(nm) dual-head YLF laser was pulsed to illuminate tracer particles. Images of the illuminated particles were acquired using two Phantom Miro Lab 320 cameras with AF Nikkor 50mm f/1.8D lenses. The cameras were situated under the test section and normal to the laser sheet, as seen in \cref{fig:Diagram}. The laser and optics were configured such that the laser sheet was concurrent with the cross-section of the cylinder. Experiments were performed using both single-pulse and double-pulse laser configurations. Single-pulse experiments produced a higher time resolution but shorter record length while double pulse experiments lead to a longer record length but reduced time resolution. The sampling rate for single pulse experiments was 70 Hz while double pulse experiments had an effective sampling rate of 12 Hz. The LaVision DaVis PIV software suite was used for calibration, image acquisition, and post-processing. The PIV algorithm used to compute velocity fields consisted of 3 passes with the final iteration using a 24-by-24 pixel window with 50\% overlap. Finally, a 3-by-3 smoothing filter was applied to reduce noise in the flow-fields.

Besides the full-field information, the velocity from an interrogation point located four diameters downstream of the cylinder will also be discussed in order to compare local and global dynamics. To maintain a constant distance from the cylinder, the interrogation point was translated in phase with the cylinder during the forcing trajectory. Welch's method was used to compute the Power Spectral Density (PSD) for the mean-subtracted, transverse component of velocity at the interrogation point. Specifically, Hamming windows with no overlap and a window size corresponding to half of the forcing cycle were used. Furthermore, each window was padded with zeros to provide a finer spectral resolution. This choice of window was used in order to identify the two dominant frequencies exhibited by the system, with one present in each half of the forcing cycle. Finite camera memory resulted in the recording of fewer cycles for cases corresponding to smaller forcing frequencies. This corresponds to the use of fewer windows in Welch's method and interrogation point spectra that may not be fully converged for lower frequency cases.

Many of the results discussed above are linked to certain phases in the cycle. Phase averaging is employed to gain a deeper understanding of these phase-locked phenomena.  Results from double-pulse experiments are used because they correspond to a longer record length and hence contain more forcing cycles. Phase-averaged results, presented in \cref{section:phaseAverage}, incorporate 42 forcing cycles with each forcing cycle consisting of 129 snapshots. Phase averaging is completed by separating each time-series into individual forcing cycles then averaging across all the cycles (corresponding to the same forcing parameters).

In addition to PIV, fluorescent dye was used to visualize unsteady structures present in the flow. A dye injection port was located at the upstream stagnation point of the cylinder and (vertically) at the center of the test section. Fluorescent water tracing dye, produced by Ecoclean Solutions, was injected at the upstream stagnation point and illuminated using two HouLight 50 W ultraviolet floodlights. A thin slit was placed in front of the UV floodlights to reduce the beam width and increase contrast in images. The flow-rate of the dye was adjusted to ensure no jet was produced (which would perturb the cylinder boundary layers), rather a thin filament of dye advected evenly on both sides of the cylinder and into the wake. A Nikon D600 DSLR camera with an AF Nikkor 50mm f/1.8D lens was used to record images of the illuminated dye. In contrast to PIV, where images were acquired in the lab-fixed frame, the DSLR was mounted onto the CTS facing down along the axis of the cylinder, i.e. images were acquired in the body-fixed frame.

\subsection{Cylinder trajectory}
\label{section:trajectory}

The cylinder was translated in the streamwise direction with harmonic variation of streamwise position while the freestream velocity was held constant. Each cylinder trajectory consisted of a sinusoidal oscillation of streamwise position with angular frequency $\omega_f=2 \pi/\tau_f$. The freestream velocity, $U_\infty$, was held constant, such that the effective instantaneous velocity experienced by the cylinder is given by
\begin{equation}
	U(t) = U_\infty -\frac{2q}{\omega_f} \sin(\omega_f t).
	\label{eqn:freestream}
\end{equation}
The parameter $q$, with units of acceleration, characterizes the amplitude of forcing, after Glaz et al. (2018) (note that the latter study was performed in a different frame of reference, namely a stationary cylinder experienced $U_\infty(t)$). The corresponding variation of instantaneous Reynolds number is:
\begin{eqnarray}
	Re(t)=\frac{U(t) D}{\nu}=Re_0 - Re_q \sin(\omega_f t).
\end{eqnarray}
Here $D$ and $\nu$ are the cylinder diameter and the kinematic viscosity of the fluid, respectively. $Re_0 = \frac{U_\infty D}{\nu}$ corresponds to the mean Reynolds number, while $Re_q = \frac{2Dq}{\nu \omega_f}$ characterizes the forced perturbation around the mean. $Re_0$ and $Re_q$, combined with the forcing Strouhal number, $St_f = \frac{\omega_f D}{2 \pi U} $, can be changed to modify the forcing trajectory. For flow over a stationary cylinder, the relation between the Strouhal number, $St_0$, and the Reynolds number, $Re_0$, is well characterized \cite{Fey_1998}. However, in order to use the framework developed above, a relation for the instantaneous shedding frequency, $f(t)$, is required.

\begin{figure}
	\captionsetup[subfigure]{oneside,margin={0.85cm,0cm}}
	\hspace{-1cm}
	\begin{subfigure}[b]{0.49\textwidth}
		\includegraphics[trim = 15px 10px 10px 210px, clip, scale=.98]{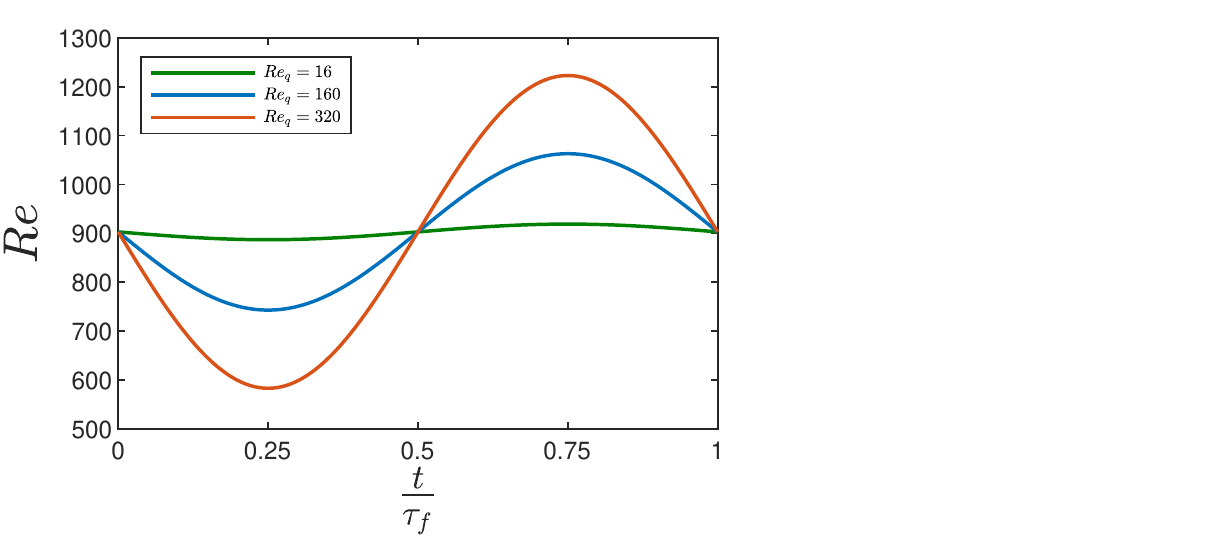}
		\caption{}
		\label{fig:traj}
	\end{subfigure}
	\begin{subfigure}[b]{0.49\textwidth}
		\hspace{0.7cm}
		\includegraphics[trim = 10px 10px 175px 210px, clip, scale=.98]{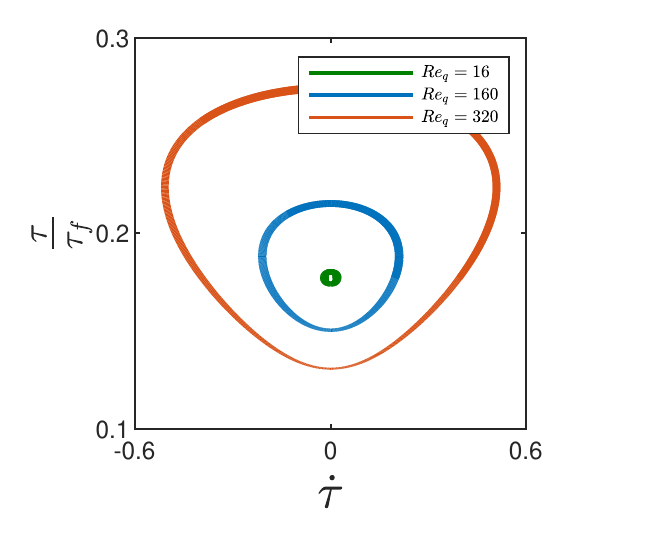}
		\caption{}
		\label{fig:phasePortrait}
	\end{subfigure}

	\captionsetup[subfigure]{oneside,margin={0.85cm,0cm}}
	\hspace{-1cm}
	\begin{subfigure}[b]{0.49\textwidth}
		\includegraphics[trim = 15px 10px 10px 210px, clip, scale=.98]{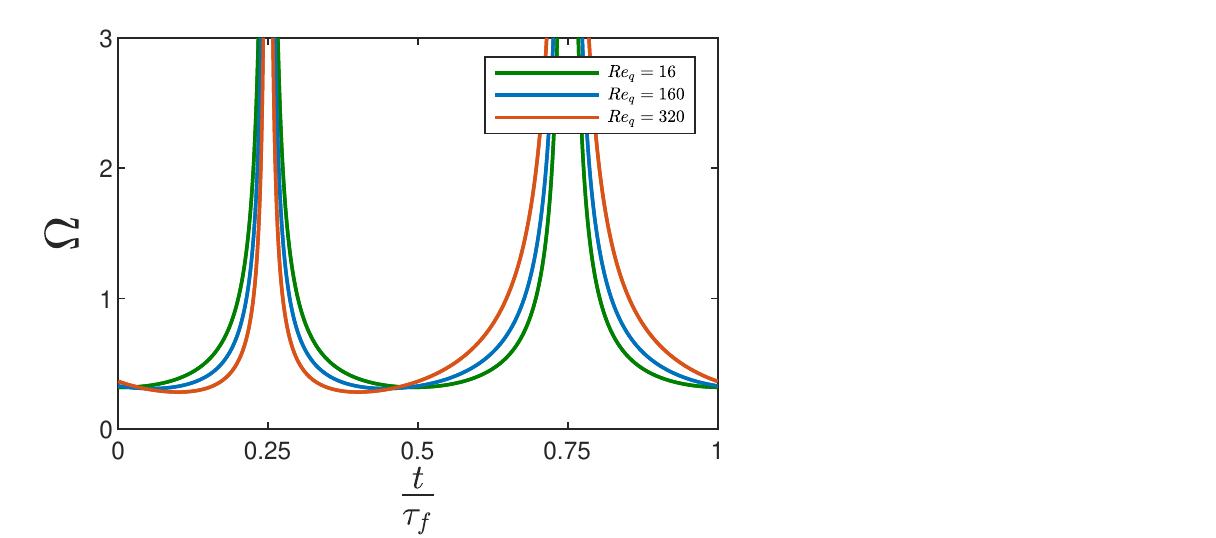}
		\caption{}
		\label{fig:quasiSteadiness}
	\end{subfigure}
	\captionsetup[subfigure]{oneside,margin={0.85cm,0cm}}
	\begin{subfigure}[b]{0.49\textwidth}
		\includegraphics[trim = 15px 10px 10px 210px, clip, scale=.98]{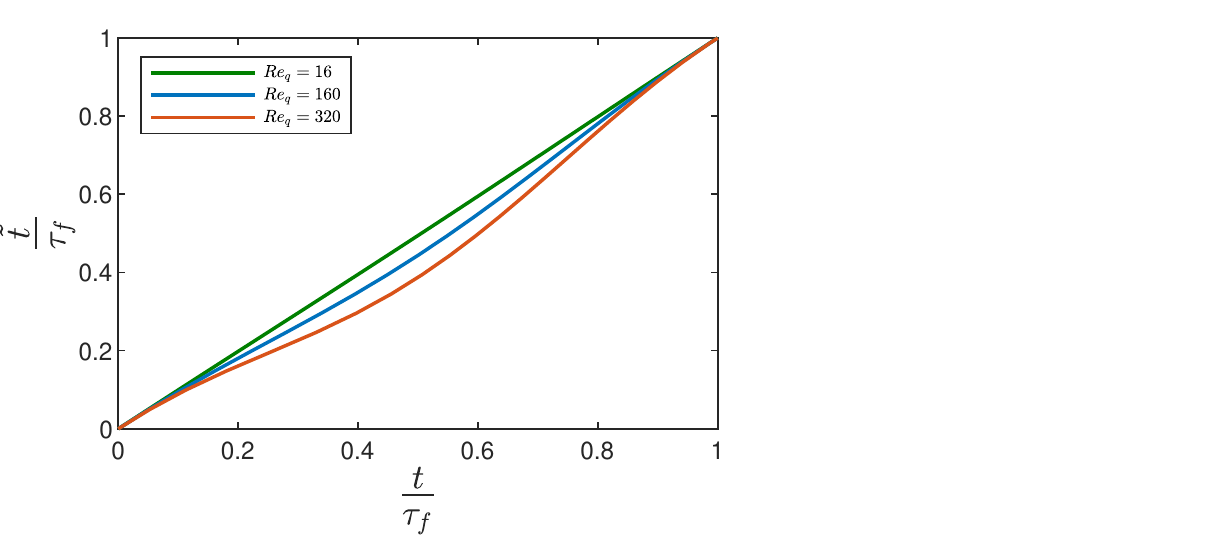}
		\caption{} 
		\label{fig:timeScaling}
	\end{subfigure}
	\captionsetup{justification=raggedright,singlelinecheck=false}
	\caption{(a) Instantaneous Reynolds number, $Re(t)$, corresponding to experimental forcing trajectory. (b) Phase portrait showing change in $\tau$ and $\dot{\tau}$ during forcing cycle. Line width is proportional to speed in phase-space. (c) Quasi-steadiness parameter plotted for streamwise forcing trajectory. (d) Variation in scaled time, $\tilde{t}$, with lab time, $t$.}
	\label{fig:scalingResults}
\end{figure}

Experiments were performed for multiple combinations of forcing frequency and amplitude at a mean Reynolds number of $Re_0 = 900$. Forcing frequencies both one and two orders of magnitude less the stationary shedding frequency were considered. Throughout the remainder of this paper, $St_f < St_0$ will refer to cases with a forcing frequency one order-of-magnitude less than  the stationary shedding frequency while $St_f \ll St_0$ will designate cases with a two order-of-magnitude difference between forcing and stationary shedding frequencies.  Table~\ref{table:testParameters} presents a summary of the various combinations of forcing frequency and amplitude considered in experiments; the time-dependent Reynolds number experienced by the cylinder is shown in \cref{fig:traj}. In this study, experiments utilized forcing parameters that ensured the instantaneous Reynolds numbers remained above the critical value at all times. It should be noted that although the mean Reynolds number considered in this study varied from that considered by Glaz et al. (2017), there was still some overlap in ratios of both forcing frequency and amplitude, $\frac{St_f}{St_0}$, and $\frac{Re_q}{Re_0} $, respectively.

\begin{table}[H]
	\centering
	\begin{tabular}{|| c | c | c | c ||}
		\hline
		$Re_0$ & $Re_q$ & $\sfrac{St_f}{St_0}$ & $ \sfrac{Re_q}{Re_0}$ \\
		\hline \hline
		900    & 16     & 0.018                & 0.018                 \\
		\hline
		900    & 160    & 0.018                & 0.18                  \\
		\hline
		900    & 16     & 0.036                & 0.018                 \\
		\hline
		900    & 160    & 0.036                & 0.18                  \\
		\hline
		900    & 320    & 0.036                & 0.35                  \\
		\hline
		900    & 16     & 0.18                 & 0.018                 \\
		\hline
		900    & 160    & 0.18                 & 0.18                  \\
		\hline
		900    & 320    & 0.18                 & 0.35                  \\
		\hline
	\end{tabular}
	\caption{Forcing parameters considered in oscillating cylinder experiments}
	\label{table:testParameters}
\end{table}

Each forcing trajectory was such that the stationary shedding Strouhal number, $St$, corresponding to the instantaneous Reynolds number was approximately constant with a value of $St\approx St_0 = 0.21$. From the definition of Strouhal number, it can be seen that for constant $St$, any change in freestream velocity, $U$, must be balanced by a variation of the shedding frequency, $f$. Making a quasi-steady assumption, namely, $St\approx St_0 = \mathrm{constant}$, allows the instantaneous shedding frequency to be expressed as
\begin{equation}
	\label{eqn:frequencyStrouhal}
	f(t)=\frac{St_0 U(t)}{D},
\end{equation}
or for the harmonic variation of cylinder velocity given in \cref{eqn:freestream},
\begin{equation}
	\label{eqn:frequency}
	f(t) = \frac{St_0 \nu}{D^2} \bigg(Re_0-Re_q \sin (\omega_f t)\bigg).
\end{equation}
The quasi-steady assumption implies that the system dynamics are such that the fundamental mechanism of vortex shedding remains the same, subject only to an adjustment in frequency in response to the change in the effective instantaneous Reynolds number. Figure~\ref{fig:phasePortrait} shows the phase portrait relating $\dot \tau$ and $\tau/\tau_f$ for three of the trajectories examined here under this assumption.

The quasi-steadiness, $\Omega(t)$ (\cref{eqn:omega}), can be computed using \cref{eqn:frequency} to give
\begin{equation}
	\label{eqn:quasisteadiness}
	\Omega = \Bigg | \frac{\sec(\omega_f t)\big(Re_0-Re_q \sin(\omega_f t)\big)^2 }{\pi\big(Re_0^2-Re_q^2\big)}\Bigg |.
\end{equation}
The time variation of $\Omega$ for various forcing regimes is shown in \cref{fig:quasiSteadiness}.
Two peaks in $\Omega$ are observed at $\frac{t}{\tau_f} = \frac{1}{4}$ and $\frac{t}{\tau_f} = \frac{3}{4}$, predicting strongly quasi-steady behavior, where the assumption that $St \approx St_0$ should hold, around those points in the forcing cycle. The characteristic frequency and Strouhal number at those two points in the cycle, which will depend on the amplitude of the cylinder trajectory through $Re_q$, can be predicted using \cref{eqn:frequency} and will be investigated in the subsequent sections.The corresponding variation of scaled time, $\tilde{t}$, shown in \cref{fig:timeScaling}, is obtained by substituting \cref{eqn:frequency} into \cref{eqn:scaling},
\begin{equation}
	\tilde{t}=t+\frac{Re_q}{Re_0 \omega_f}\bigg[\cos(\omega_f t)-1\bigg].
	\label{eqn:scaledTime}
\end{equation}

\subsection{Koopman Mode Decomposition \& Time Scaling}\
Koopman Analysis was used to extract spatio-temporally significant flow structures and gain a deeper understanding of the physical mechanisms underlying the flow. The Koopman modes and eigenvalues were approximated using Dynamic Mode Decomposition (DMD) \cite{Rowley_2009}, \cite{Schmid_2010}. The specific algorithm used to decompose velocity fields is detailed in Tu et al. (2014)\cite{Tu_2014}.

\begin{figure}[H]
	\begin{center}
		\includegraphics[trim =0px 0px 0px 0px,clip]{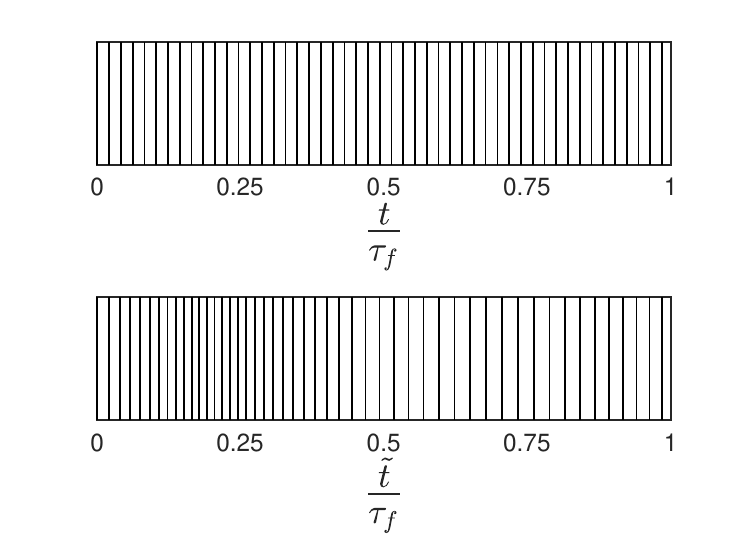}
	\end{center}
	\caption{Experimental sampling rate in lab time (top) and corresponding scaled time (bottom). Vertical lines correspond to snapshot acquisition in the experiment. For the sake of clarity, the number of samples shown has been significantly decreased relative to the actual experiment.}
	\label{fig:sampling}
\end{figure}

KMD was first performed on PIV velocity fields acquired using a constant sampling rate relative to lab time, $t$, i.e. with a constant temporal separation between snapshots. Next KMD was performed on PIV velocity fields corresponding to a constant sampling rate relative to scaled time, $\tilde{t}$. The former analysis is straightforward.  In the latter case, PIV snapshots do not correspond to a constant sampling rate with respect to $\tilde{t}$. The sampling rate used for PIV is shown relative to both lab time, $t$, and scaled time, $\tilde{t}$, in \cref{fig:sampling}. However, the mapping between $t$ and $\tilde{t}$, specifically \cref{eqn:scaledTime}, is known, thus the time-series were temporally interpolated to generate time-series that correspond to constant sampling frequencies with respect to $\tilde{t}$. Interpolated time-series will be referred to as a scaled data set in the remainder of this work. It was observed that time scaling could lead to numerical instabilities in the corresponding dataset's KMD. Reasons for this will be elaborated on in \cref{section:Discussion} but these instabilities were remedied by performing KMD using every second snapshot in the time-series. Although this resulted in a lower effective sampling frequency, the effective sampling rate used for KMD was approximately 70 times higher than the stationary shedding frequency. Furthermore, KMD of unscaled datasets using all the snapshots showed little difference when compared to cases using every second snapshot for KMD. Unscaled datasets had no issues with numerical stability but KMD was still performed using every second snapshot in the time-series in order to facilitate comparison with scaled data.

Phase averaging results in a time-series corresponding to a single (phase averaged) forcing cycle. Although the temporal resolution of the time-series does not change, the total length of the time-series is reduced. As a result, KMD extracts a smaller number of modes and eigenvalues. KMD of phase averaged velocity fields incorporated all the snapshots in the time-series in order to increase the temporal resolution and number of resulting modes and eigenvalues. Even though double-pulse PIV velocity fields were used for phase averaging, the effective KMD sampling rate was still approximately 25 times higher than the stationary shedding frequency.

KMD applied to a PIV time-series yields both Koopman modes and eigenvalues. Eigenvalues describe the temporal behavior (i.e., growth-rate and frequency) of the spatial structure in the associated mode. In what follows, the spectral amplitudes for each Koopman mode consist of the average projection coefficient of all the snapshots in the time-series onto the respective mode. As will be shown in subsequent sections, the transverse velocity field, $v$, serves as a good indicator of vortex shedding. Given that the shedding frequency is taken to be the characteristic frequency of the system, KMD is performed on the transverse velocity field. Vorticity is also a good indicator of shedding but computing gradients using PIV velocity fields can generate significant noise which may corrupt the KMD. Although data acquisition took place in the lab frame, relative to which the cylinder was moving, Koopman analysis is performed in the cylinder-fixed frame. All of the flow structures present in the flow are generated by the cylinder, hence the cylinder-fixed frame is the natural choice for analysis.

\section{Results}
\label{section:Results}
\subsection{Stationary cylinder}
\begin{figure}
	\begin{subfigure}[b]{0.3\textwidth}
		\begin{center}
			\includegraphics[trim = 20px 140px 55px 120px,clip,scale=0.4]{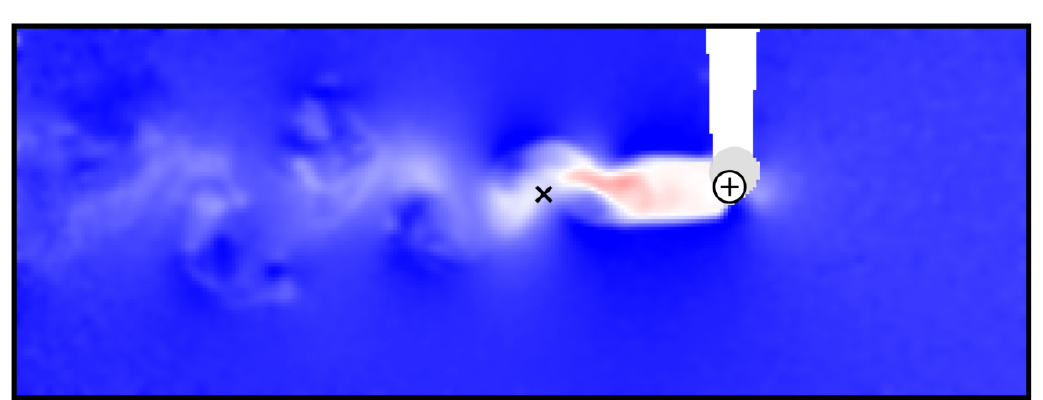}
			\caption{}
		\end{center}
	\end{subfigure}
	\begin{subfigure}[b]{0.3\textwidth}
		\begin{center}
			\includegraphics[trim = 20px 140px 55px 120px,clip,scale=0.4]{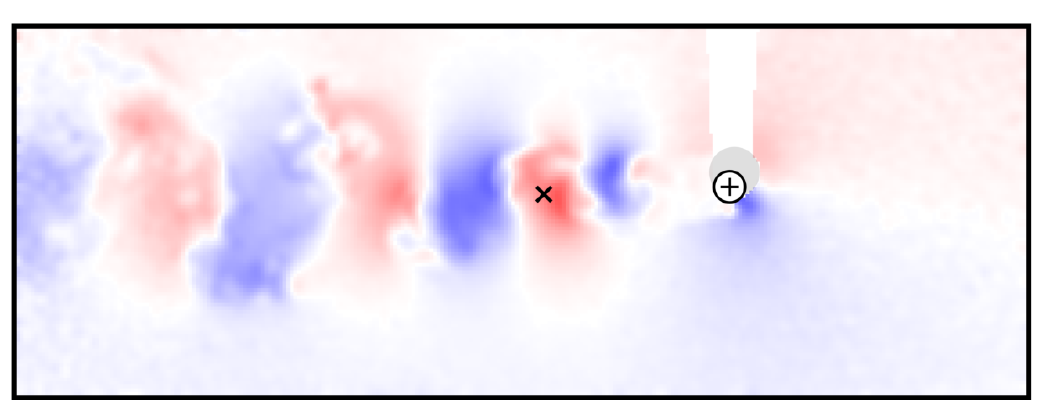}
			\caption{}
			\label{fig:stationaryV}
		\end{center}
	\end{subfigure}
	\begin{subfigure}[b]{0.3\textwidth}
		\begin{center}
			\includegraphics[trim = 20px 140px 55px 120px,clip,scale=0.4]{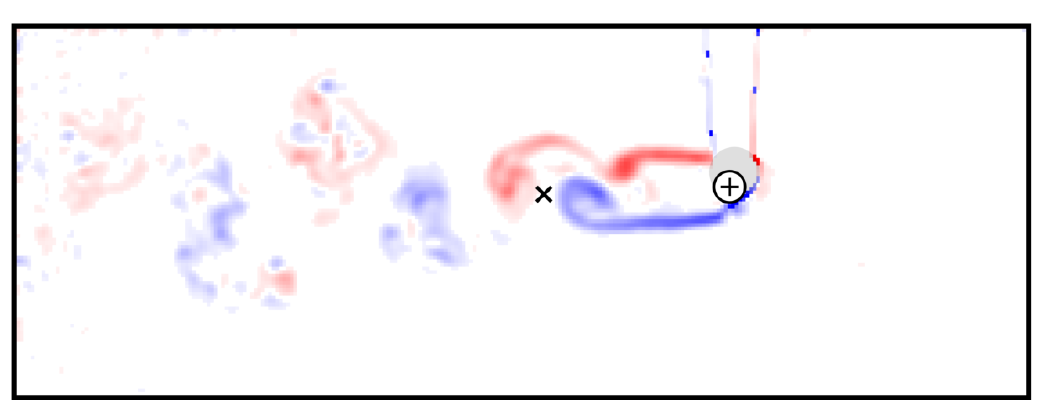}
			\caption{}
		\end{center}
	\end{subfigure}
	\captionsetup{justification=raggedright,singlelinecheck=false}
	\caption{Snapshots of streamwise velocity (a), transverse velocity (b), and vorticity (c) for the stationary cylinder. The cross-section of the cylinder at the measurement plane is represented by $\oplus$ and the shaded circle corresponds to an obstruction in the field of view due to the bottom of the cylinder. The location of the interrogation point is denoted by $\boldsymbol{\times}$. Flow is from right to left.}
	\label{fig:stationaryWake}
\end{figure}

\begin{figure}
	\begin{subfigure}[b]{1\textwidth}
		\includegraphics[trim = 170px 25px 0 180px, clip,scale = 0.63]{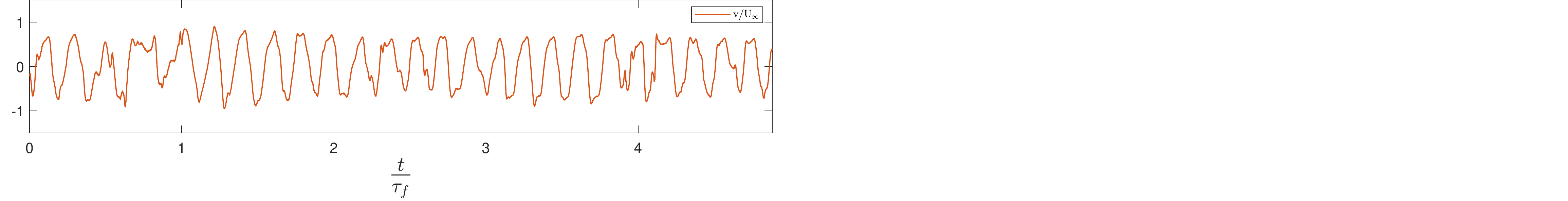}
		\label{fig:stationaryTimeTrace}
	\end{subfigure}
	\begin{subfigure}[b]{1\textwidth}
		\includegraphics[trim = 170px 25px 0 180px, clip,scale = 0.63]{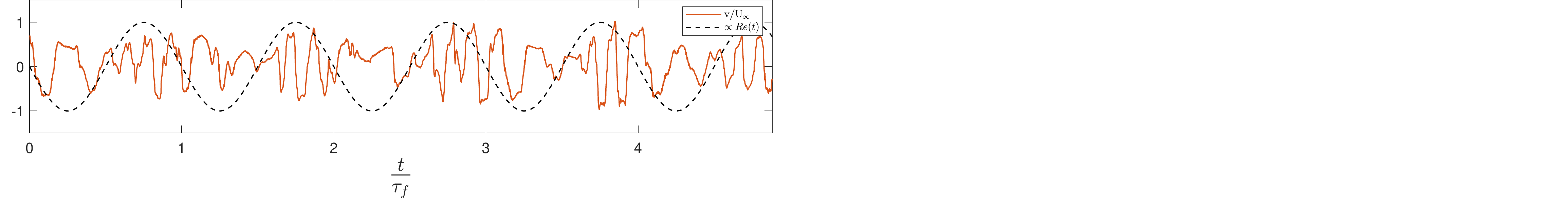}
		\label{fig:forcedTimeTraceFast}
	\end{subfigure}
	\begin{subfigure}[b]{1\textwidth}
		\includegraphics[trim = 170px 25px 0 180px, clip,scale = 0.63]{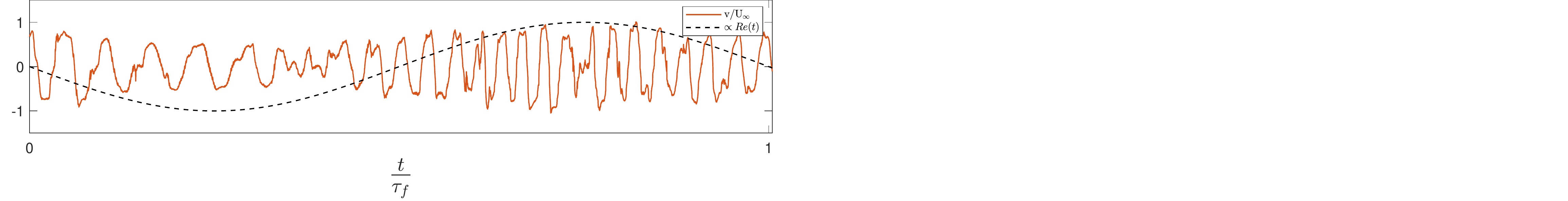}
		\label{fig:forcedTimeTraceSlow}
	\end{subfigure}
	\captionsetup{justification=raggedright,singlelinecheck=false}
	\caption{Time-trace of transverse velocity, $\mathrm{v}/\mathrm{U}_\infty$, at interrogation point. The dashed line is proportional to the instantaneous Reynolds number, $Re(t)$. Top: stationary cylinder; middle:  $Re_0=900,$ $\sfrac{Re_q}{Re_0} = 0.35,$ $\sfrac{St_f}{St_0}=0.18$; bottom: $Re_0=900,$ $\sfrac{Re_q}{Re_0} = 0.35,$ $\sfrac{St_f}{St_0}=0.036$.}
	\label{fig:timeTrace}
\end{figure}
\begin{figure}
	\captionsetup[subfigure]{oneside,margin={0.35cm,0cm}}
	\begin{subfigure}[b]{0.39\textwidth}
		\includegraphics[trim = 35px 15px 145px 190px,clip,scale = 0.87]{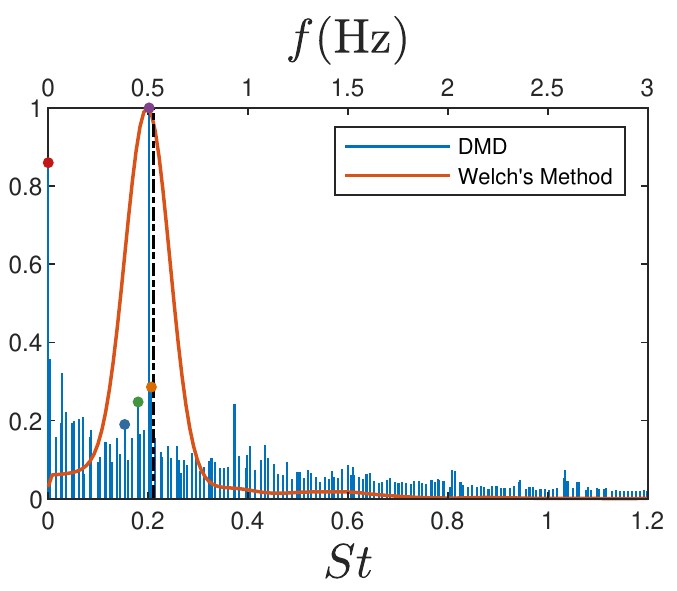}
		\caption{}
		\label{fig:KMD_Spectrum_Stationary}
	\end{subfigure}
	\captionsetup[subfigure]{oneside,margin={-0.05cm,0cm}}
	\begin{subfigure}[b]{0.6\textwidth}
		\includegraphics[trim = 25px 0px 25px 0px,clip, scale = 1.35]{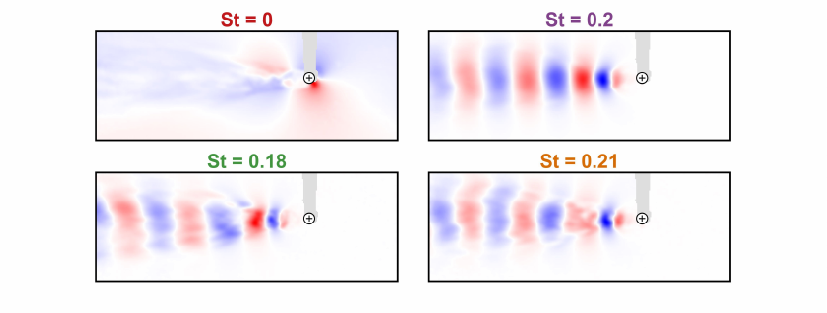}
		\vspace{0.4cm}
		\caption{}
		\label{fig:KMD_Modes_Stationary}
	\end{subfigure}
	\captionsetup{justification=raggedright,singlelinecheck=false}
	\caption{KMD spectrum (a) and modes (b) for stationary cylinder, $Re_0 = 900$. With the exception of the mean mode ($St = 0$), the colored circles on the spectrum represent shedding modes. The colored mode labels correspond to the circles of the same color on the spectrum. The black dot-dashed line represents the stationary shedding frequency.}
	\label{fig:KMD_Stationary}
\end{figure}

Results for the stationary cylinder are reviewed briefly to provide the reference configuration. The instantaneous PIV velocity and vorticity fields in \cref{fig:stationaryWake}  correspond to a classic Karman vortex street. As seen in \cref{fig:timeTrace}, the transverse component of velocity, $v$, measured at the interrogation point demonstrates an approximately constant dominant frequency, as vortices of opposite sense of rotation convect past. The power-spectral density of the transverse velocity at the interrogation point, \cref{fig:KMD_Spectrum_Stationary}, confirms that the shedding Strouhal number was almost identical to that predicted in the literature for this Reynolds number, $St\approx 0.21$ \cite{Fey_1998}.

\begin{figure}
	\includegraphics[trim = 24px 20px 200px 200px,clip,scale=1.2]{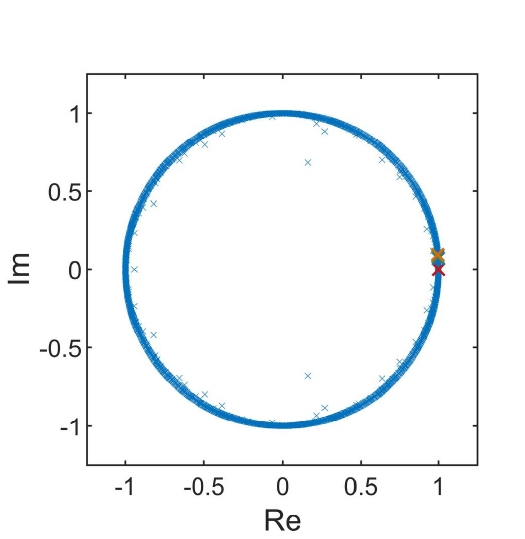}
	\caption{Koopman eigenvalues for stationary cylinder, $Re_0=900$. The colored symbols correspond to the modes indicated in \cref{fig:KMD_Stationary}.}
	\label{fig:KMD_Eigenvalues_Stationary}
\end{figure}

A plot of the Koopman eigenvalues for the stationary cylinder (\Cref{fig:KMD_Eigenvalues_Stationary}) indicates that all eigenvalues lie on or inside the unit circle, such that the corresponding Koopman modes either decay with time or oscillate periodically. This is true for the forced cases as well, so, for brevity, plots of eigenvalues are omitted for the rest of the cases discussed. The KMD spectrum for the stationary case is shown in \cref{fig:KMD_Spectrum_Stationary}, allowing a comparison of local (interrogation point) analysis and global (KMD) {decomposition}. Both methods have a peak near $St \approx 0.2$, while the KMD spectrum for the stationary cylinder has additional, low level amplitude over a range of Strouhal numbers around the shedding peak.
Comparing the KMD modes shown in \cref{fig:KMD_Modes_Stationary} to the transverse component of velocity measured using PIV (\cref{fig:stationaryV}), the similarity between vertical structures in both the transverse velocity field and KMD modes is evident. Furthermore, the modes shown, which are identified by the colored markers on the KMD spectrum, are clustered around the shedding frequency, with the exception of the mean mode, which reflects the anticipated flow deflection around the cylinder. These KMD modes, which will henceforth be referred to as ``shedding modes", are flow structures associated with vortex shedding. The finite width of the spectral peak and distribution of multiple shedding modes around it in \cref{fig:KMD_Spectrum_Stationary} indicates that the shedding in the stationary case might not be purely periodic. Rather, shedding is limited to a small band of frequencies around the expected stationary value. This can likely be attributed to small fluctuations in the pump frequency driving the flow in the water channel as well as secondary instabilities present in this flow regime. It should be noted that the shedding mode (\cref{fig:KMD_Modes_Stationary}) corresponding to the peak in the KMD spectrum, very close to the predicted shedding frequency, has the cleanest spatial shape, with minimal noise and contamination from other structures present in the flow.

\subsection{Oscillating cylinder}
\subsubsection{Koopman mode analysis of the wake}
Under streamwise oscillation of the cylinder, a range of flow phenomena not present in the stationary case develop. For brevity, results are presented for two combinations of forcing parameters but similar results results were observed for the other forcing trajectories as well. Specifically, results are presented for one amplitude ratio, $\sfrac{Re_q}{Re_0}=0.35$, and two forcing frequency ratios, $\sfrac{St_f}{St_0} = 0.18 $ and $\sfrac{St_f}{St_0} = 0.036$.

\begin{turnpage}
	\begin{figure}
		\begin{subfigure}[b]{0.31\textwidth}
			\begin{center}
				\hspace*{0.25cm}\includegraphics[trim =0 0 0 0, clip,height =1cm,frame=0.4mm,left]{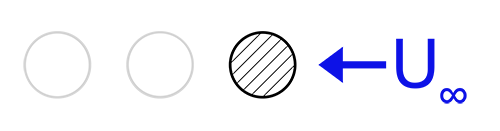}
			\end{center}
		\end{subfigure}
		\begin{subfigure}[b]{0.31\textwidth}
			\begin{center}
				\hspace*{0.25cm}\includegraphics[trim =0 0 0 0, clip,height =1cm,frame=0.4mm,left]{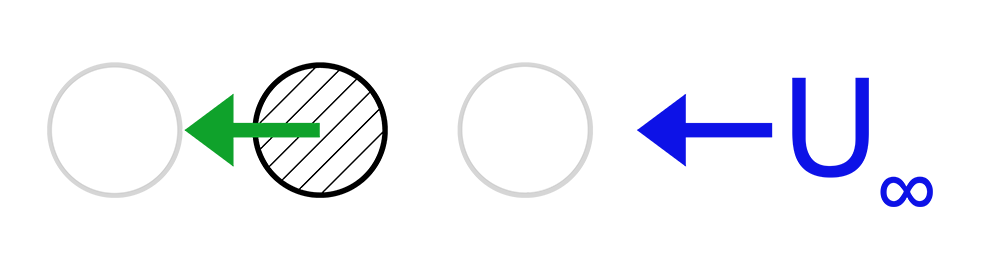}
			\end{center}
		\end{subfigure}
		\begin{subfigure}[b]{0.31\textwidth}
			\begin{center}
				\hspace*{0.25cm}\includegraphics[trim =0 0 0 0, clip,height =1cm,frame=0.4mm,left]{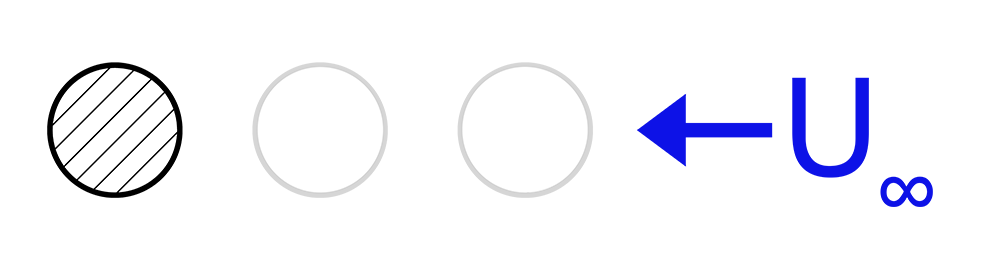}
			\end{center}
		\end{subfigure}
		\begin{subfigure}[b]{0.31\textwidth}
			\begin{center}
				\hspace*{0.25cm}\includegraphics[trim =0 0 0 0, clip,height =1cm,frame=0.4mm,left]{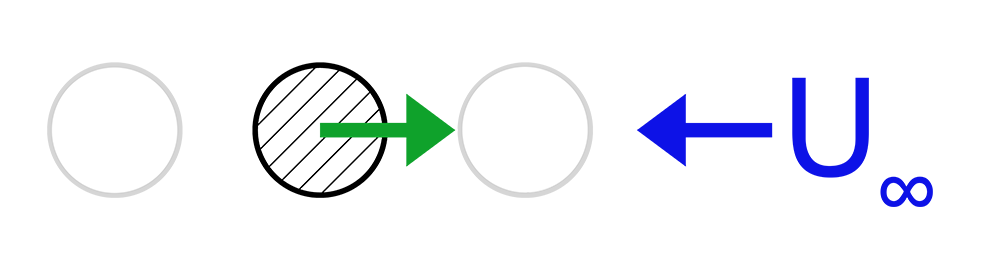}
			\end{center}
		\end{subfigure}

		\vspace{0.6cm}
		\begin{subfigure}[b]{0.31\textwidth}
			\begin{center}
				\includegraphics[trim =15.8cm 2.6cm 8cm 17.5cm, clip,scale=0.1,left]{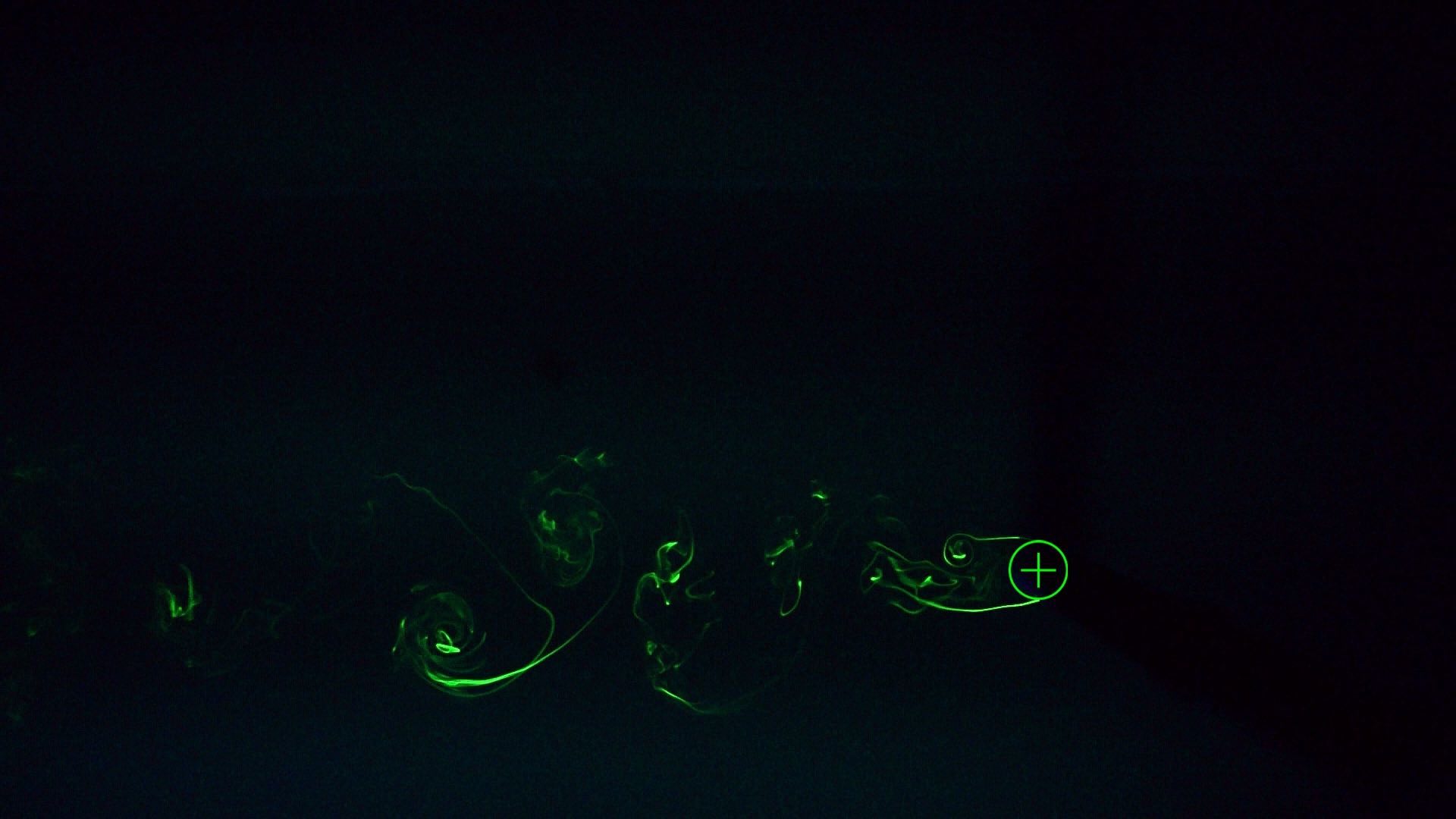}
			\end{center}
		\end{subfigure}
		\begin{subfigure}[b]{0.31\textwidth}
			\begin{center}
				\includegraphics[trim =15.8cm 2.6cm 8cm 17.5cm, clip,scale=0.1,left]{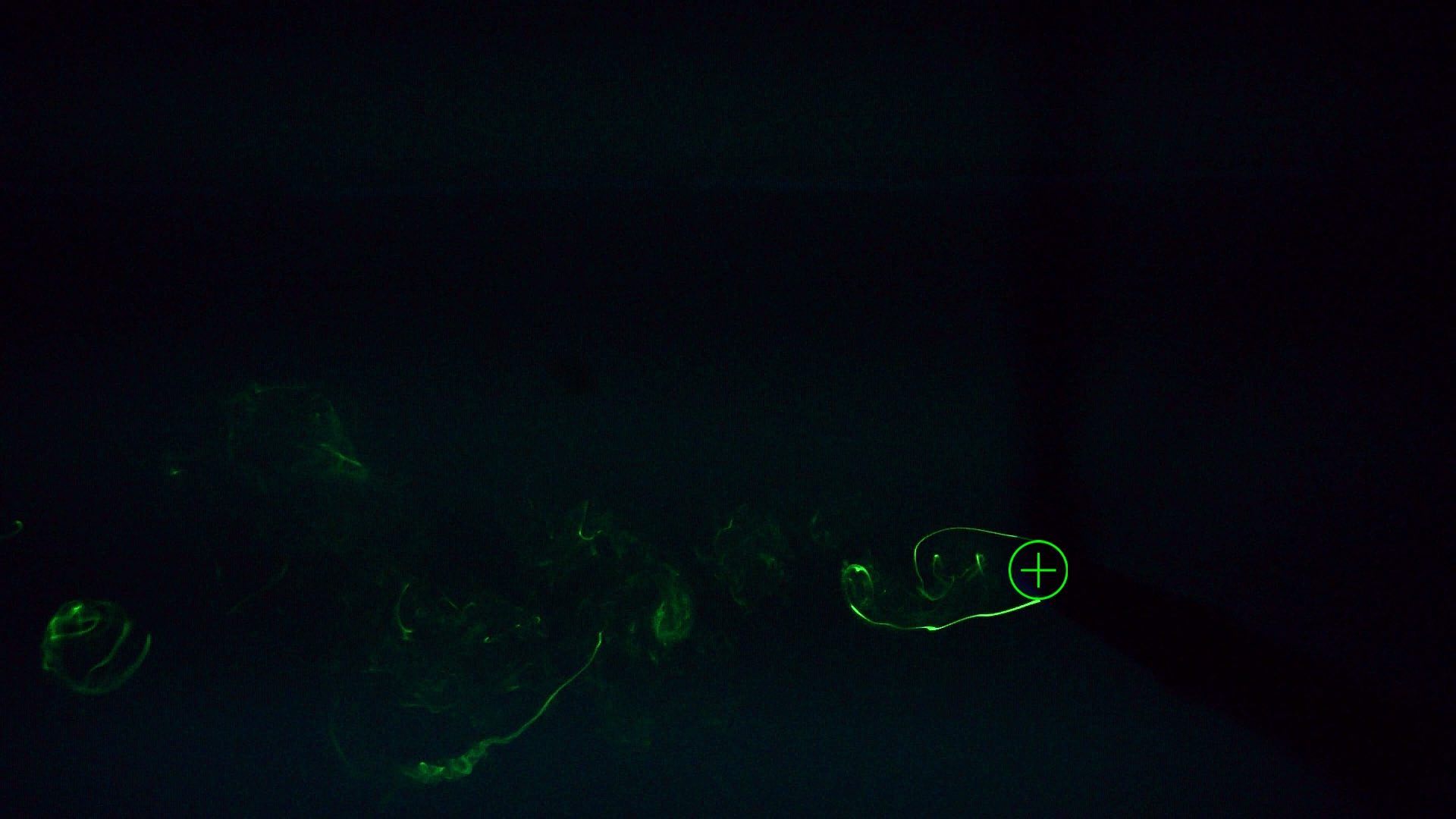}
			\end{center}
		\end{subfigure}
		\begin{subfigure}[b]{0.31\textwidth}
			\begin{center}
				\includegraphics[trim =15.8cm 2.6cm 8cm 17.5cm, clip,scale=0.1,left]{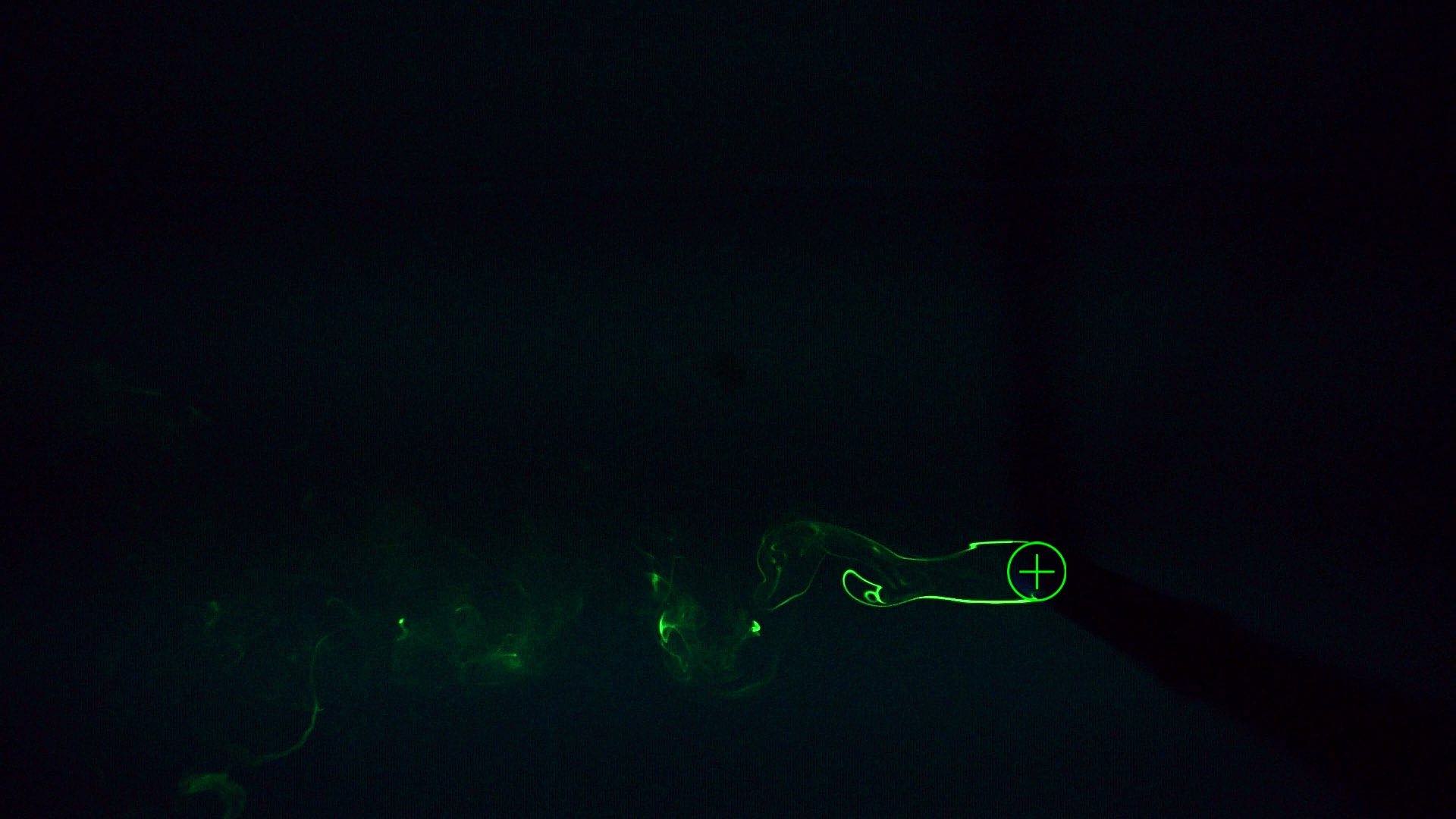}
			\end{center}
		\end{subfigure}
		\begin{subfigure}[b]{0.31\textwidth}
			\begin{center}
				\includegraphics[trim =16cm 2.6cm 7.75cm 17.5cm, clip,scale=0.1,left]{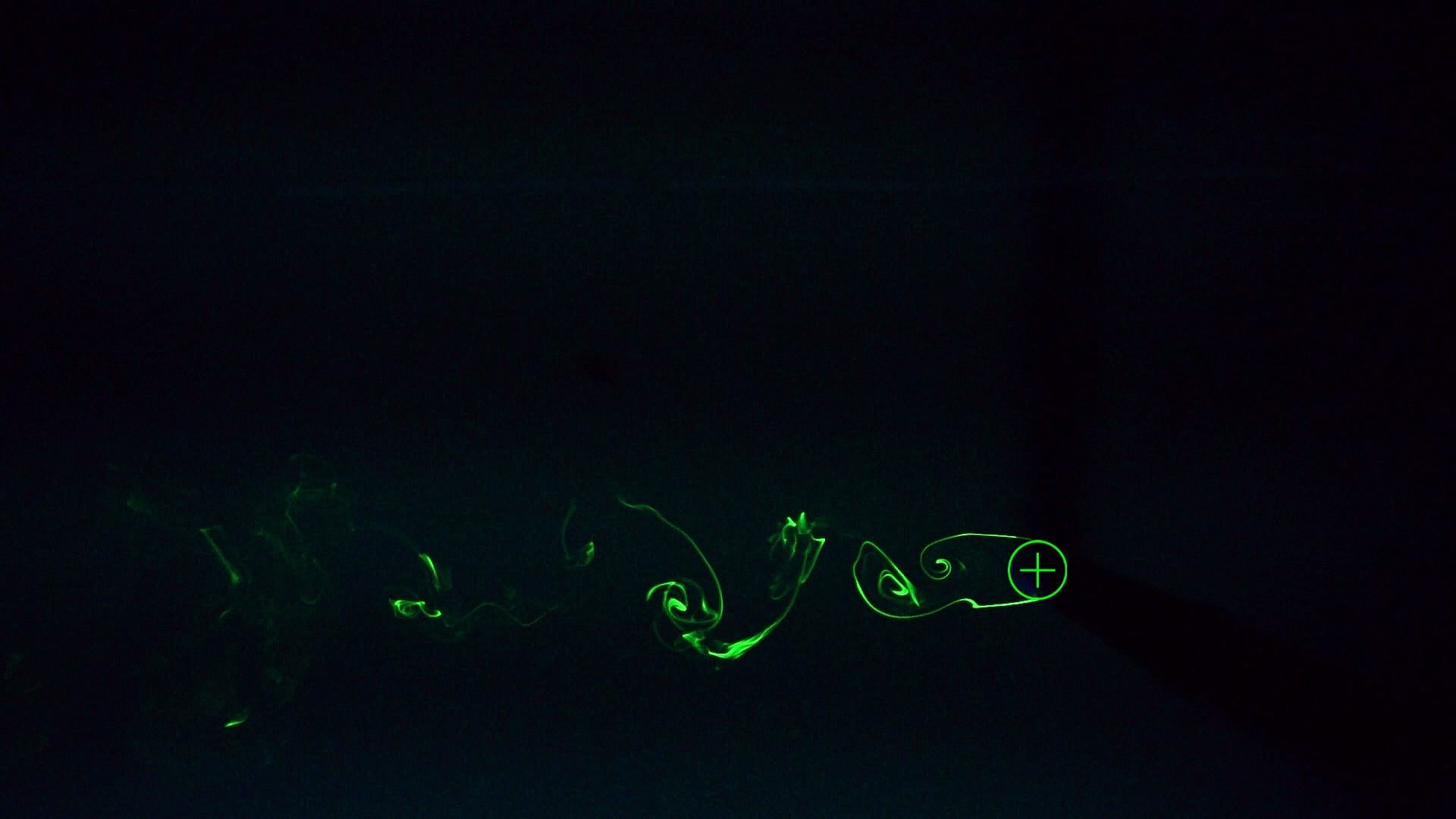}
			\end{center}
		\end{subfigure}

		\vspace{0.6cm}
		\begin{subfigure}[b]{0.31\textwidth}
			\begin{center}
				\includegraphics[trim =2.1cm 1.3cm 3.1cm 1cm, clip,height=2cm,left]{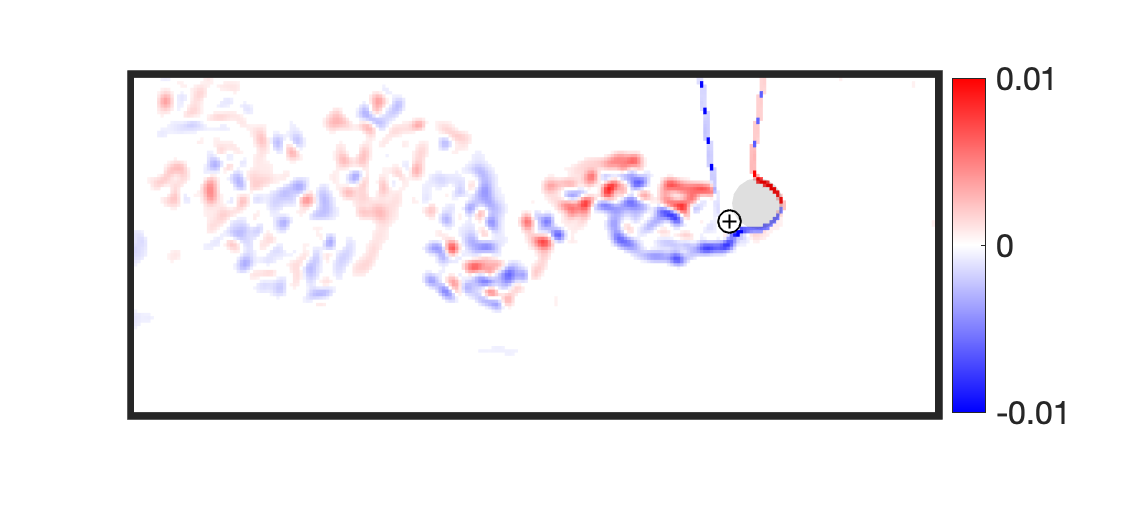}
			\end{center}
		\end{subfigure}
		\begin{subfigure}[b]{0.31\textwidth}
			\begin{center}
				\includegraphics[trim =2.1cm 1.3cm 3.1cm 1cm, clip,height=2.0cm,left]{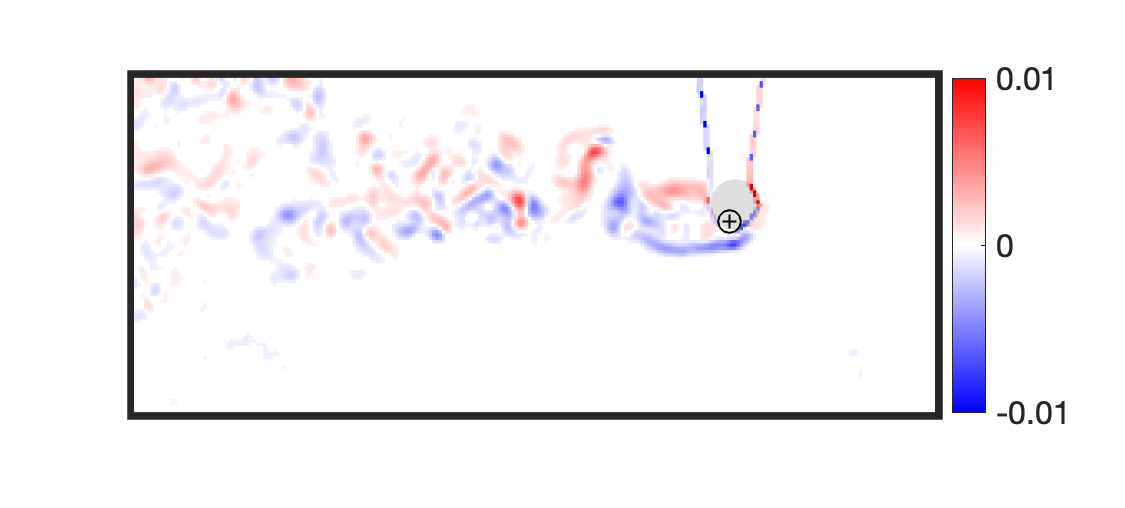}
			\end{center}
		\end{subfigure}
		\begin{subfigure}[b]{0.31\textwidth}
			\begin{center}
				\includegraphics[trim =2.1cm 1.3cm 3.1cm 1cm, clip,height=2.0cm,left]{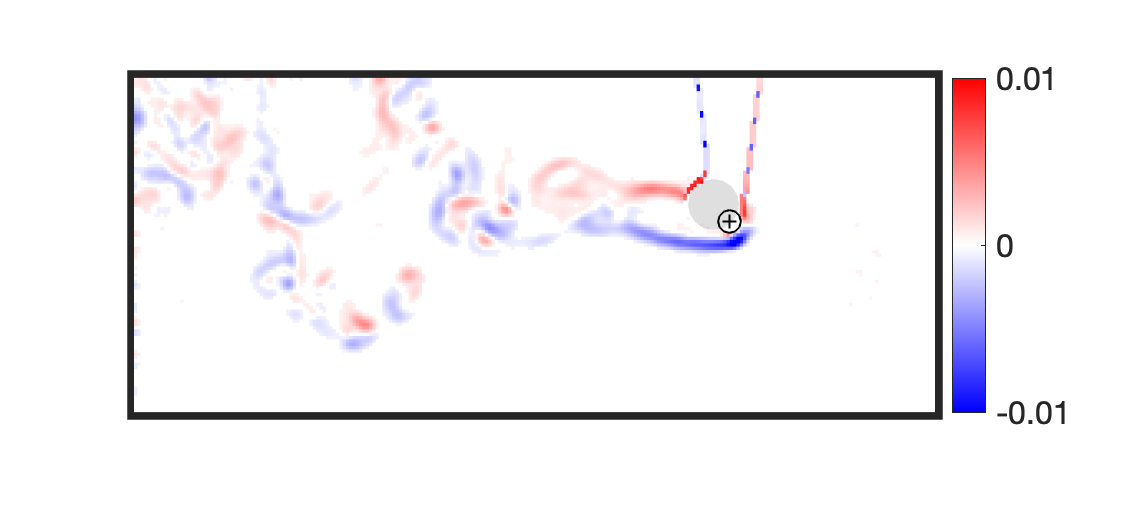}
			\end{center}
		\end{subfigure}
		\begin{subfigure}[b]{0.31\textwidth}
			\begin{center}
				\includegraphics[trim =2.1cm 1.3cm 0.9cm 1cm, clip,height=2.0cm,left]{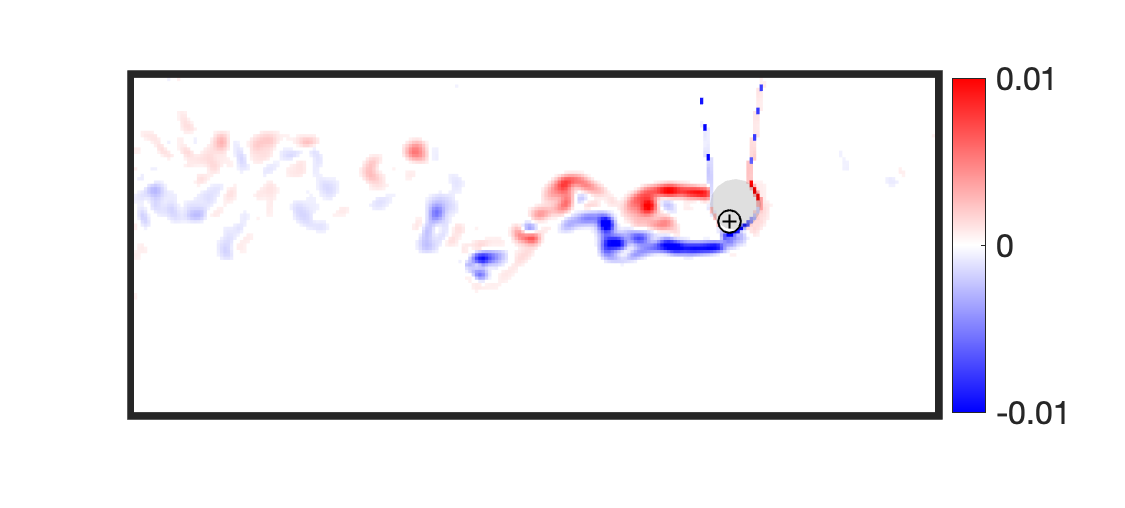}
			\end{center}
		\end{subfigure}

		\vspace{0.6cm}
		\begin{subfigure}[b]{0.31\textwidth}
			\begin{center}
				\includegraphics[trim =2.1cm 1.3cm 3.1cm 1cm, clip,height=2.0cm,left]{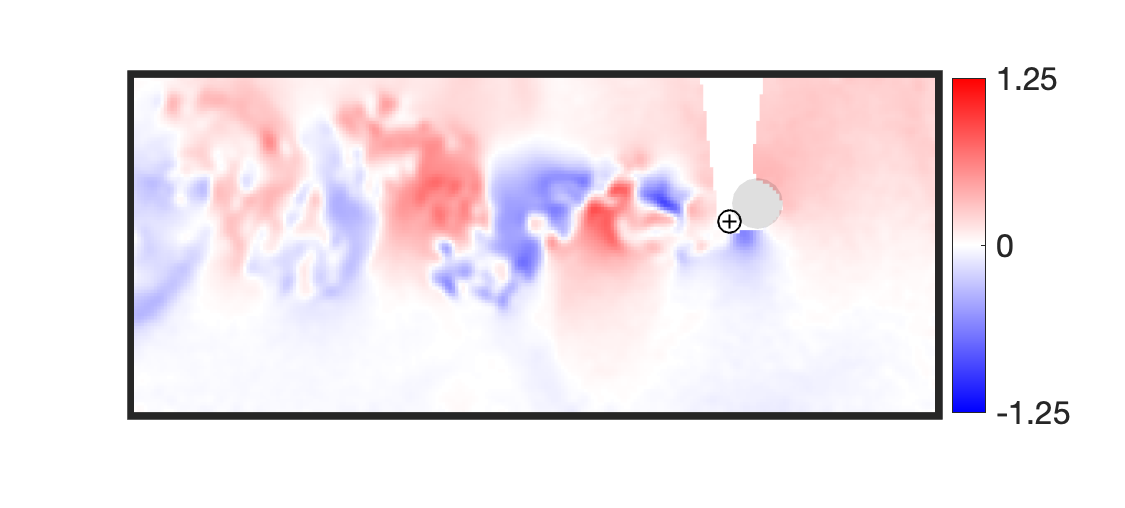}
			\end{center}
		\end{subfigure}
		\begin{subfigure}[b]{0.31\textwidth}
			\begin{center}
				\includegraphics[trim =2.1cm 1.3cm 3.1cm 1cm, clip,height=2.0cm,left]{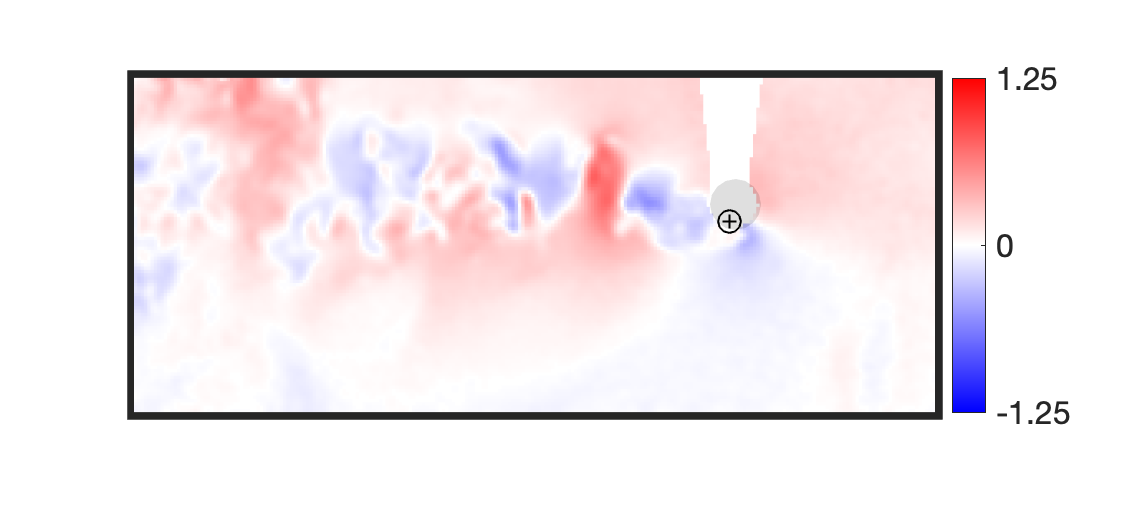}
			\end{center}
		\end{subfigure}
		\begin{subfigure}[b]{0.31\textwidth}
			\begin{center}
				\includegraphics[trim =2.1cm 1.3cm 3.1cm 1cm, clip,height=2.0cm,left]{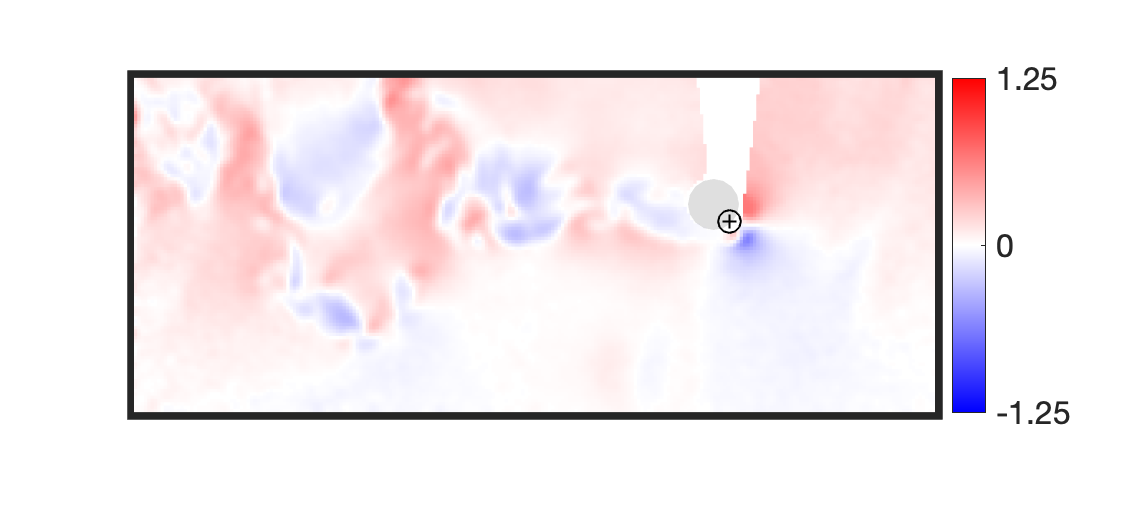}
			\end{center}
		\end{subfigure}
		\begin{subfigure}[b]{0.31\textwidth}
			\begin{center}
				\includegraphics[trim =2.1cm 1.3cm 0.9cm 1cm, clip,height=2.0cm,left]{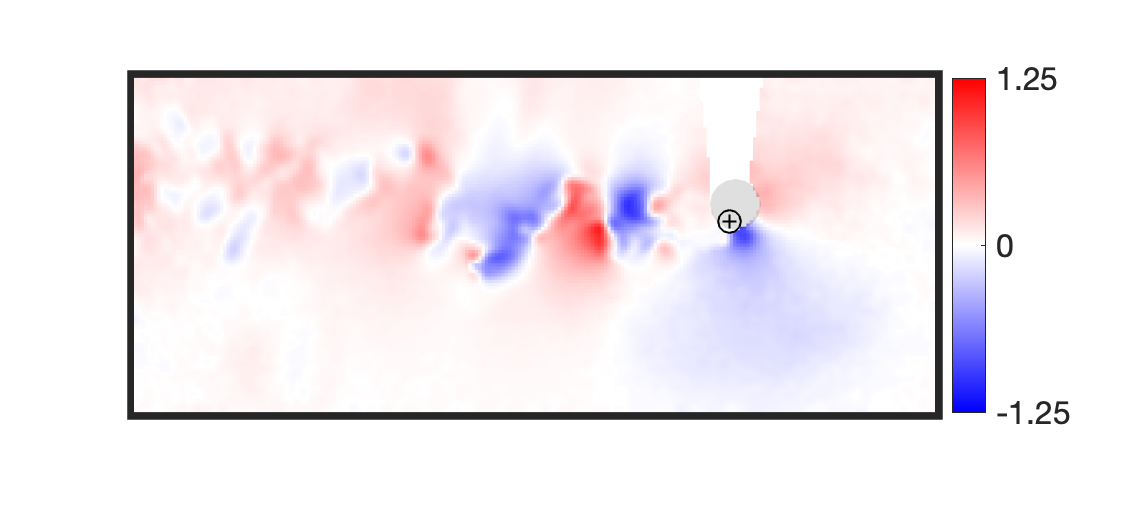}
			\end{center}
		\end{subfigure}

		\vspace{0.6cm}
		\begin{subfigure}[b]{0.31\textwidth}
			\begin{center}
				\includegraphics[trim = 1.03cm 0.6cm 2.7cm 0.4cm, clip,height =2cm,left]{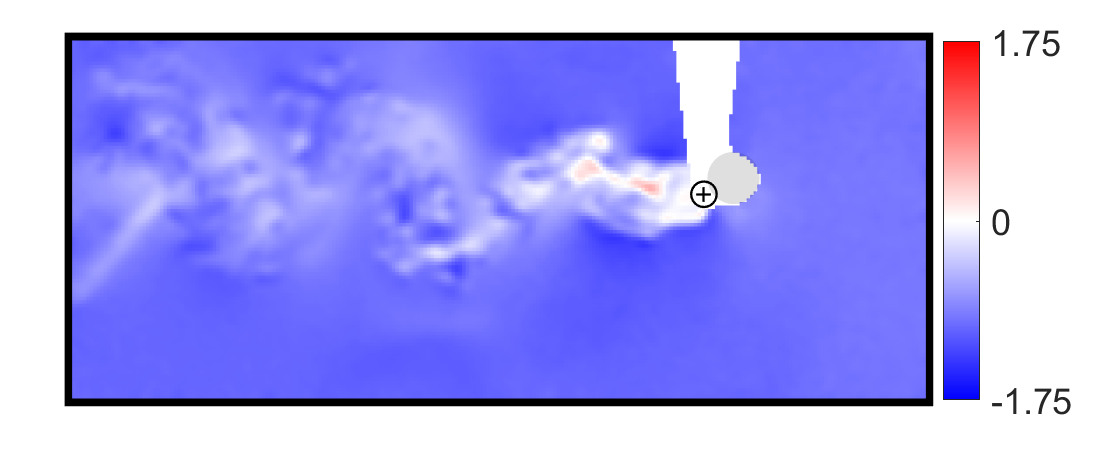}
			\end{center}
		\end{subfigure}
		\begin{subfigure}[b]{0.31\textwidth}
			\begin{center}
				\includegraphics[trim = 1.03cm 0.6cm 2.7cm 0.4cm, clip,height =2cm,left]{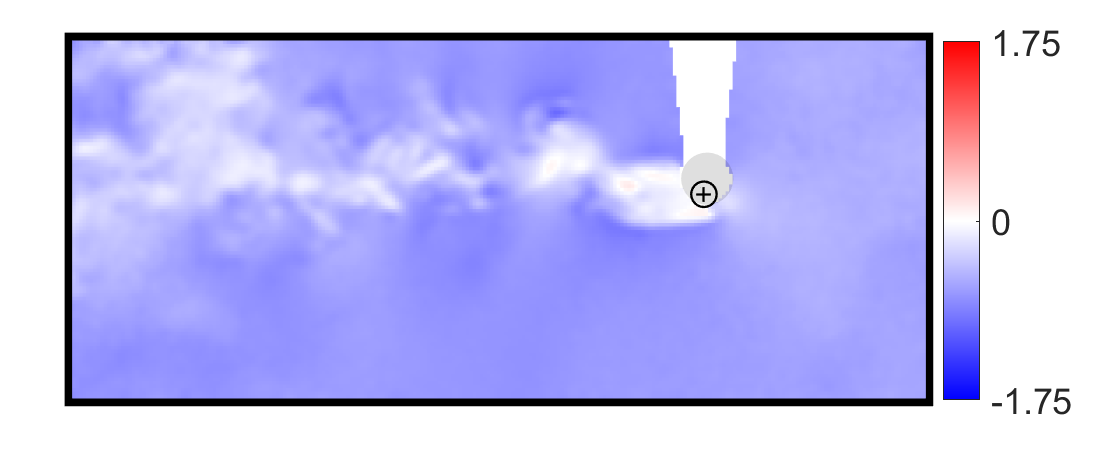}
			\end{center}
		\end{subfigure}
		\begin{subfigure}[b]{0.31\textwidth}
			\begin{center}
				\includegraphics[trim = 1.03cm 0.6cm 2.7cm 0.4cm, clip,height =2cm,left]{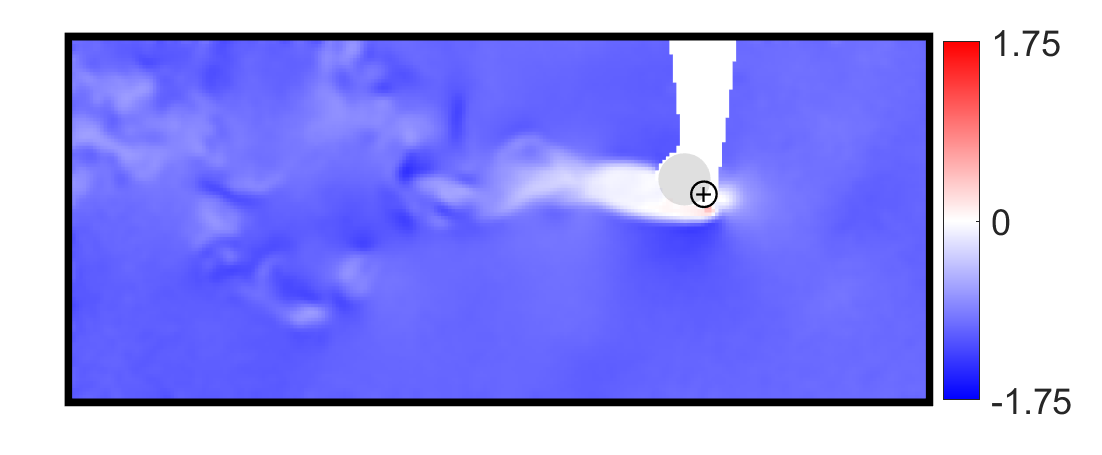}
			\end{center}
		\end{subfigure}
		\begin{subfigure}[b]{0.31\textwidth}
			\begin{center}
				\includegraphics[trim = 1.03cm 0.6cm 0cm 0.4cm, clip,height =2cm,left]{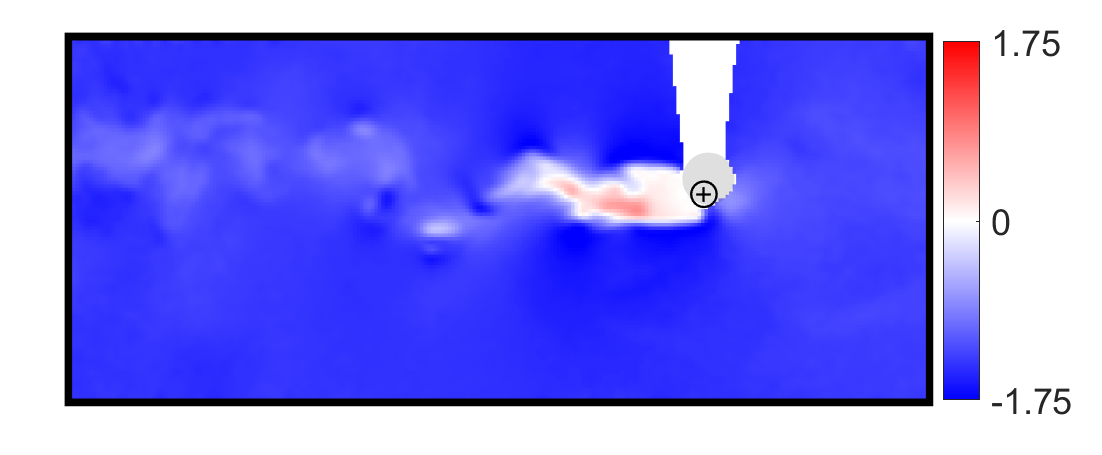}
			\end{center}
		\end{subfigure}
			\captionsetup{justification=raggedright,singlelinecheck=false}
		\caption{Flow around a streamwise oscillating cylinder, $Re_0=900, \sfrac{Re_q}{Re_0} = 0.35, \sfrac{St_f}{St_0}=0.18$. The top row corresponds to the position of the cylinder in the forcing cycle. The second row corresponds to visualization of the flow using fluorescent dye. The third row corresponds to vorticity $\omega_z$. The fourth and fifth rows correspond to the transverse and streamwise components of velocity, $\frac{\textrm{v}}{\textrm{U}_\infty}$ and $\frac{\textrm{u}}{\textrm{U}_\infty}$, respectively. From left to right, the columns correspond to $\sfrac{t}{\tau_f} = 0, 0.25, 0.50$, and $0.75$, respectively. The $\bigoplus$ symbol represents the cross-section of the cylinder at the measurement plane while the shaded circle represents an obstruction in the field of view due to the bottom of the cylinder. Flow is from right to left.}
		\label{fig:composite}
	\end{figure}
\end{turnpage}

Snapshots of dye-flow visualization and PIV are shown for different phases of the forcing cycle in \cref{fig:composite}. The most notable effect of forcing on the wake was the spatial and temporal variation of the wake during the forcing cycle, observable as frequency and amplitude modulation of each component of velocity and out-of-plane vorticity.

The columns in \cref{fig:composite} correspond to four different phases in the forcing cycle. From left to right, the columns correspond to $t/\tau_f = 0, 0.25, 0.50,$ and $0.75$, respectively. At $t/\tau_f = 0$, the wake resembles that of the stationary cylinder with shed vortices of opposite sign convecting downstream. Although the larger scale structure of the wake is similar to the stationary case, streamwise forcing also generates smaller scale structures and unsteadiness which can be seen in the velocity and vorticity fields. The cylinder reaches maximum downstream velocity when $t/\tau_f = 0.25$. It can be seen that the strength of shed vortices and the corresponding shear-layers have decreased at this point in the forcing cycle. Although the instantaneous freestream velocity seen by the cylinder at $t/\tau_f = 0$ and $0.50$ is identical, the wake at the upstream turnaround point is not identical to its downstream counterpart. In the latter case, the wake no longer resembles that of the stationary cylinder. At $t/\tau_f=0.50$, the wake is relatively quiescent and the near wake sees the development of two symmetric shear-layers which roll up into a pair of symmetric vortices. The behavior at the downstream turnaround point will be discussed in more detail in \cref{section:phaseAverage}. At $t/\tau_f = 0.75$, the cylinder attains its maximum upstream velocity and also reaches its most active state. At this point in the forcing trajectory vortex shedding has resumed and the increase in instantaneous freestream velocity seen by the cylinder leads to stronger shear layers and shed vortices. The modulation in strength of shed vortices is also observed at the interrogation point in the form of the amplitude modulation, as seen in \cref{fig:timeTrace}.

More detail can be observed in the time traces at the interrogation point, shown in \cref{fig:timeTrace}. The amplitude modulation was stronger for cases corresponding to $St_f<St_0$ while the frequency modulation was more pronounced for cases where $St_f\ll St_0$. It is important to note, however, that both forcing frequencies still lead to amplitude and frequency modulation, to varying extents and correlated with the instantaneous Reynolds number seen by the cylinder, shown schematically by the dashed line as a reference to when various flow phenomena occur relative to the phase of the forcing cycle. It can be seen that shedding at points corresponding to $t\approx \frac{1}{4} \tau_f+k \tau_f$ and $t\approx \frac{3}{4} \tau_f+k \tau_f$, where $k = 0,1,2,3,...$, was the most consistent. That is, shedding occurred at a roughly constant frequency at those points in the forcing cycle. Those points in the cycle also correspond to peaks in $\Omega$ where strong quasi-steady behavior is expected, see \cref{fig:quasiSteadiness}.

\begin{figure}
	\captionsetup[subfigure]{oneside,margin={0.3cm,0cm}}
	\begin{subfigure}[b]{0.39\textwidth}
		\includegraphics[trim = 35px 15px 145px 190px,clip,scale = 0.87]{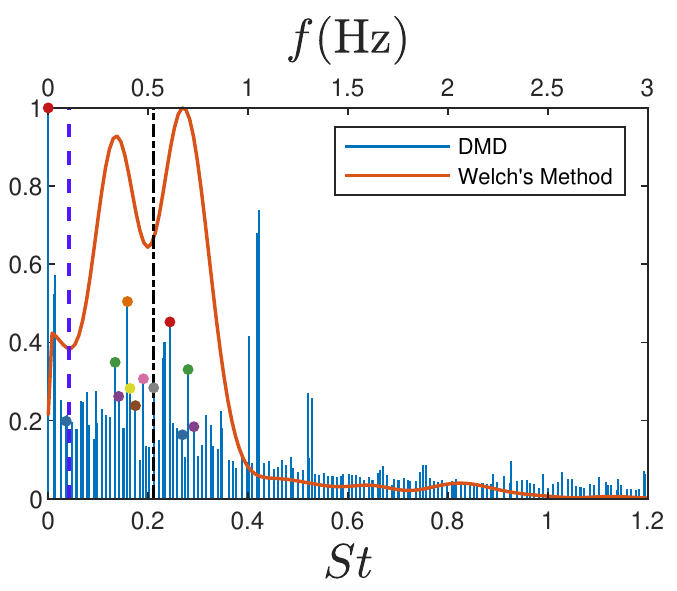}
		\caption{}
		\label{fig:forcedKMDSpectrumFast}
	\end{subfigure}
	\captionsetup[subfigure]{oneside,margin={-0.1cm,0cm}}
	\begin{subfigure}[b]{0.55\textwidth}
		\includegraphics[trim = 20px 0 20px 0px,clip, scale = 1.35]{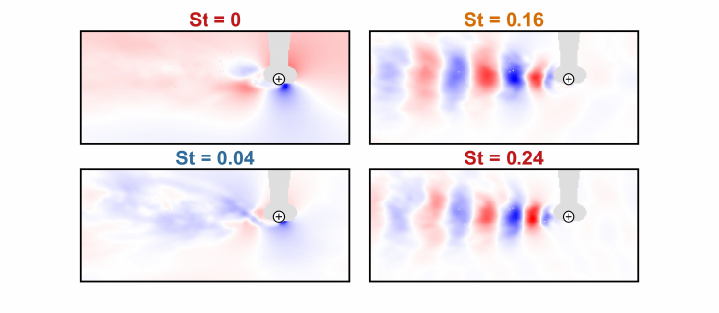}
		\vspace{0.4cm}
		\caption{}
		\label{fig:forcedKMDModesFast}
	\end{subfigure}
	\captionsetup{justification=raggedright,singlelinecheck=false}
	\caption{KMD spectrum (a) and modes (b) for oscillating cylinder, $Re_0=900,$ $\sfrac{Re_q}{Re_0} = 0.35,$ $\sfrac{St_f}{St_0}=0.18$. The colored mode labels correspond to the circles of the same color on the spectrum. The black dot-dashed line represents the stationary shedding frequency while the purple dashed line denotes the forcing frequency. The additional colored symbols on the spectrum indicate shedding modes.}
	\label{fig:forcedKMDFast}
\end{figure}

\begin{figure}
	\captionsetup[subfigure]{oneside,margin={0.3cm,0cm}}
	\begin{subfigure}[b]{0.39\textwidth}
		\includegraphics[trim = 35px 15px 145px 190px,clip,scale = 0.87]{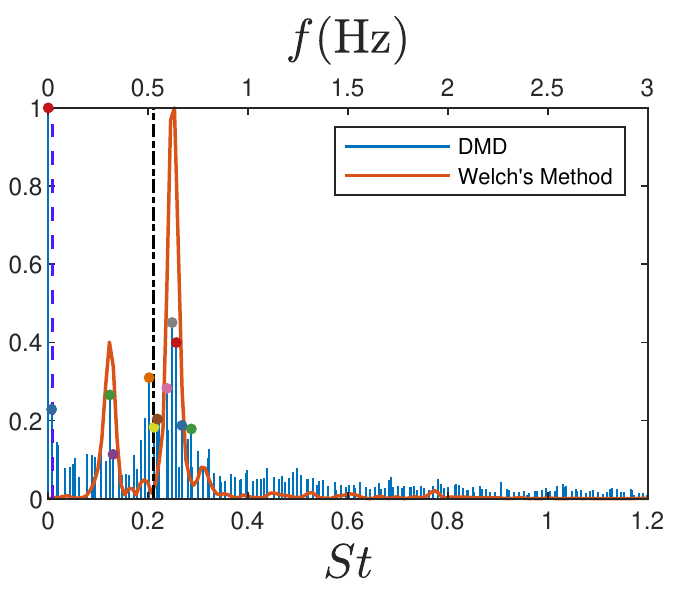}
		\caption{}
		\label{fig:forcedKMDSpectrumSlow}
	\end{subfigure}
	\captionsetup[subfigure]{oneside,margin={-0.1cm,0cm}}
	\begin{subfigure}[b]{0.55\textwidth}
		\includegraphics[trim = 0px 0 0px 0px,clip, scale = 1.35]{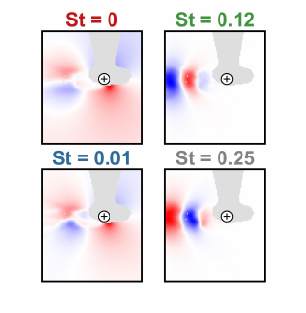}
		\vspace{0.4cm}
		\caption{}
		\label{fig:forcedKMDModesSlow}
	\end{subfigure}
	\captionsetup{justification=raggedright,singlelinecheck=false}
	\caption{KMD spectrum (a) and modes (b) for oscillating cylinder, $Re_0=900,$ $\sfrac{Re_q}{Re_0} = 0.35,$ $\sfrac{St_f}{St_0}=0.036$.}
\end{figure}

The PSDs for both forcing regimes (\cref{fig:forcedKMDSpectrumFast,fig:forcedKMDSpectrumSlow}) are broader than the stationary case, reflecting the enhanced vortical activity in the wake, with the development of two distinct peaks especially clear for the lower forcing frequency. Changes are also observed in the KMD spectra. When $Re_q$ is fixed but forcing frequency is decreased by an order of magnitude, additional shedding modes are extracted from KMD of the PIV time-series. These additional shedding modes arise because of the gradual modulation of shedding frequency during a forcing period. Four representative KMD modes are shown in \cref{fig:forcedKMDModesFast,fig:forcedKMDModesSlow} for each case, selected to reflect the mean, a mode close to the forcing frequency, and the strongest shedding modes in each spectral peak. In each case, the latter two panels show similar features to the stationary cylinder shedding mode, but at a modified frequency. A similar phenomena was presented by Arbabi \& Mezic (2017) who observed the persistence of structurally similar Koopman modes in lid driven cavity flow over a wide range of Reynolds numbers.  The zero frequency ``mean modes" also resemble their stationary counterpart in that they contain an anti-symmetric structure in the near wake. Relative to the stationary case, however, streamwise forcing generates small variations in strength and structure of the mean modes.

It was also observed that the peak in the KMD spectrum widened with forcing amplitude (not shown), a phenomenon also seen by Glaz et al. (2017). Considering cases with different forcing amplitudes, $Re_q$, but all other parameters held constant, it is observed that the separation between spectral peaks increases with forcing amplitude for both the interrogation point and KMD spectra.

\subsubsection{Structure generated at the forcing frequency}
\label{section:phaseAverage}

Besides the wake structure discussed above, the forcing frequency itself imposes temporal structure on the flow-field. This can be recovered by phase-averaging the flow fields, shown here at a series of phases during the forcing cycle for $Re_q/Re_0 = 0.35$ and $St_f/St_0 = 0.18$.

\begin{figure}
	\hspace{-0.75cm}
	\begin{subfigure}[b]{0.23\textwidth}
		\includegraphics[trim = 20px 130px 55px 120px,clip,scale=0.325]{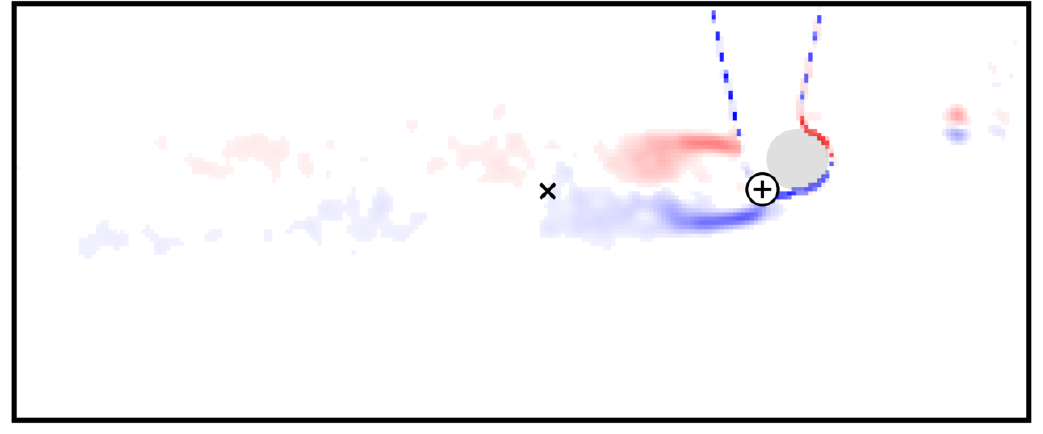}
	\end{subfigure}
	\begin{subfigure}[b]{0.23\textwidth}
		\includegraphics[trim = 20px 130px 55px 120px,clip,scale=0.325]{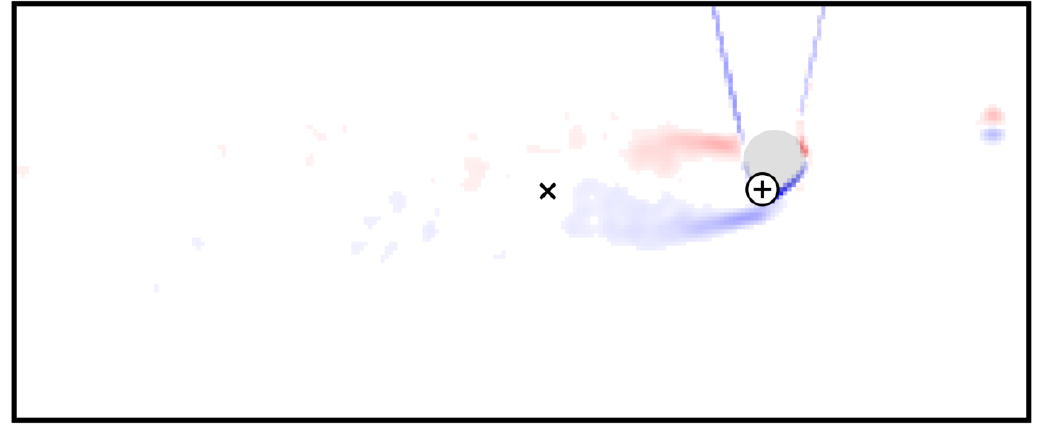}
	\end{subfigure}
	\begin{subfigure}[b]{0.23\textwidth}
		\includegraphics[trim = 20px 130px 55px 120px,clip,scale=0.325]{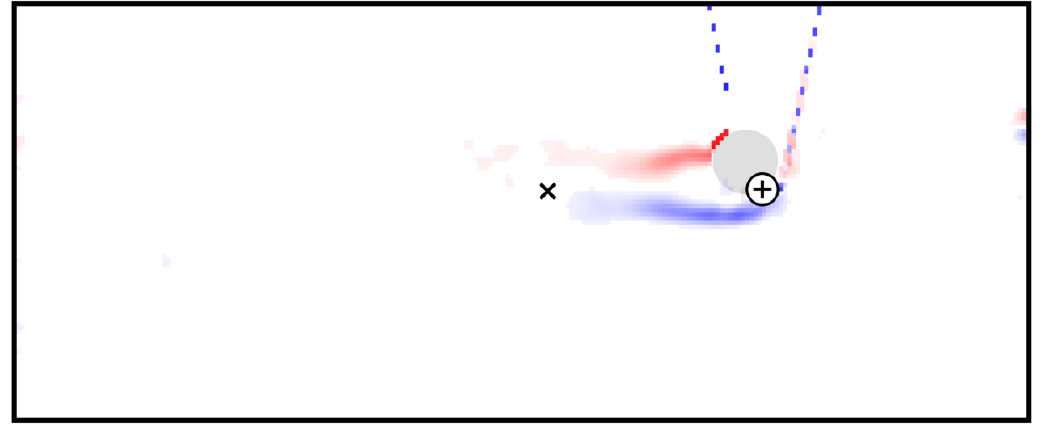}
	\end{subfigure}
	\begin{subfigure}[b]{0.24\textwidth}
		\includegraphics[trim = 20px 130px 20px 120px,clip,scale=0.325]{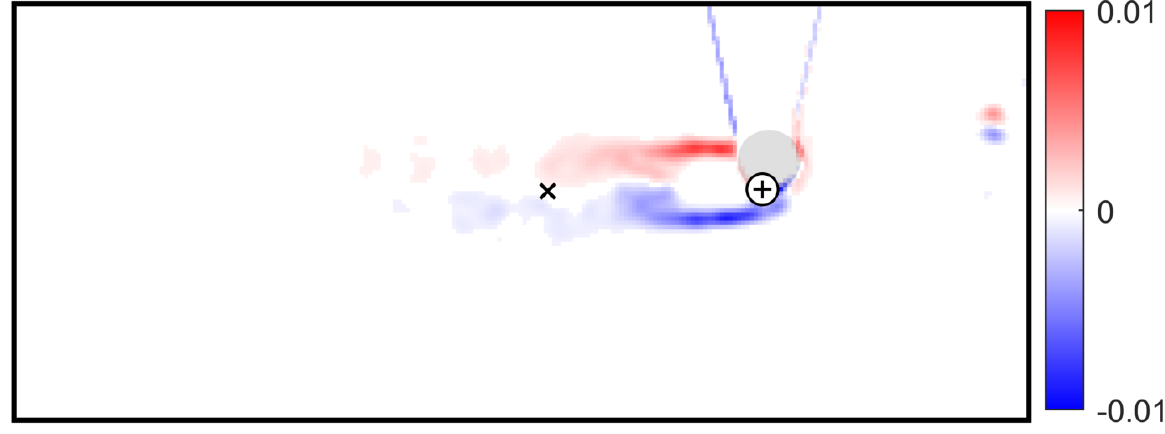}
	\end{subfigure}

	\vspace{4px}
	\hspace{-0.75cm}
	\begin{subfigure}[b]{0.23\textwidth}
		\includegraphics[trim = 20px 130px 55px 120px,clip,scale=0.325]{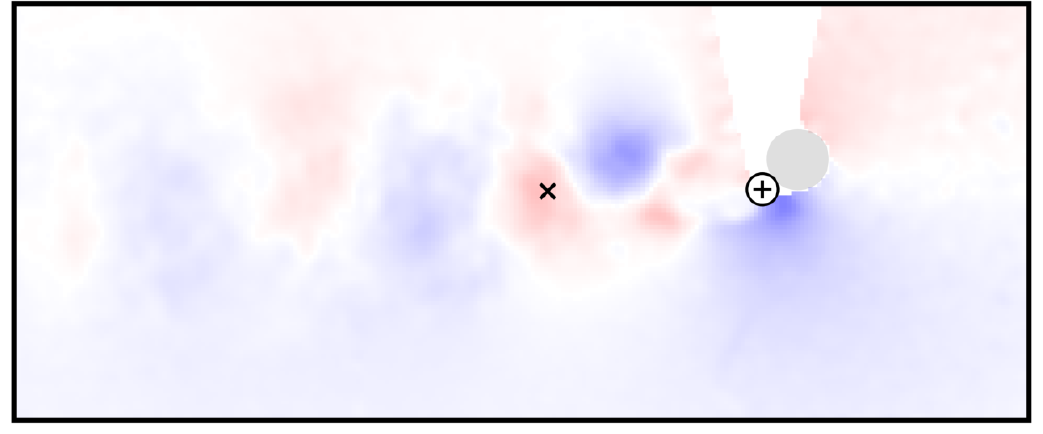}
		\caption{}
	\end{subfigure}
	\begin{subfigure}[b]{0.23\textwidth}
		\includegraphics[trim = 20px 130px 55px 120px,clip,scale=0.325]{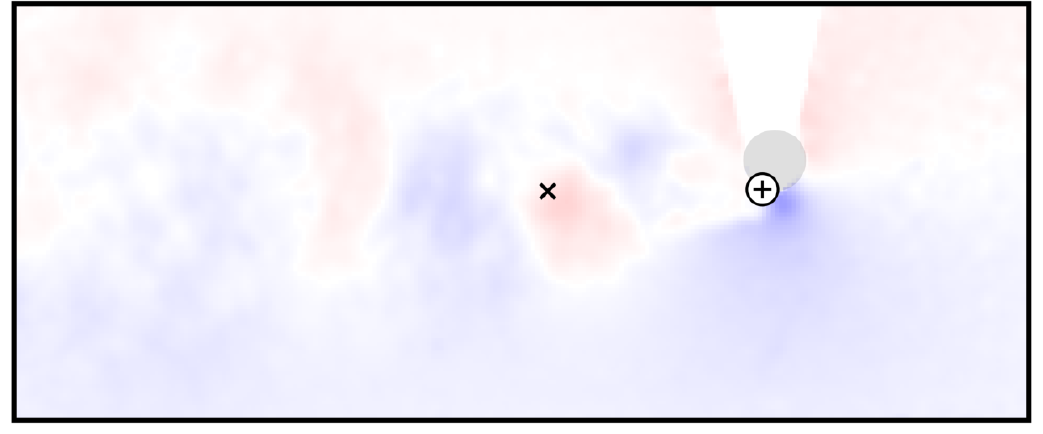}
		\caption{}
	\end{subfigure}
	\begin{subfigure}[b]{0.23\textwidth}
		\includegraphics[trim = 20px 130px 55px 120px,clip,scale=0.325]{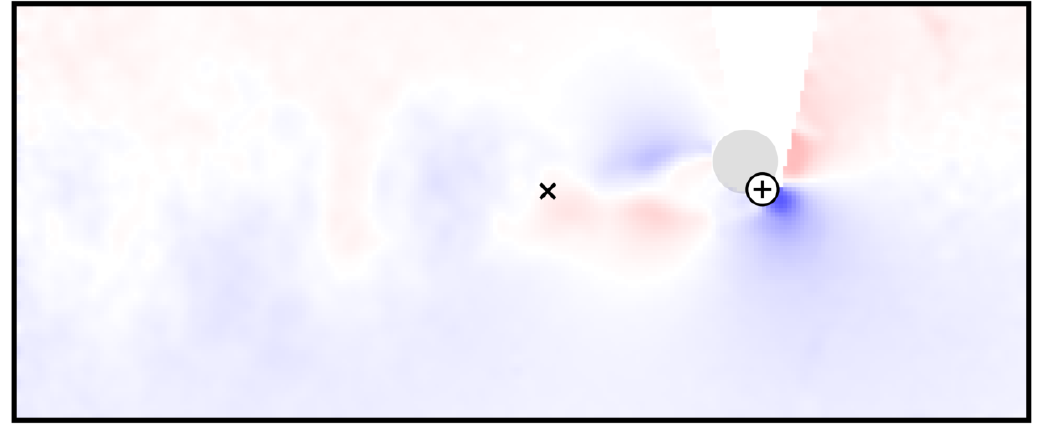}
		\caption{}
	\end{subfigure}
	\begin{subfigure}[b]{0.24\textwidth}
		\includegraphics[trim = 20px 130px 20px 120px,clip,scale=0.325]{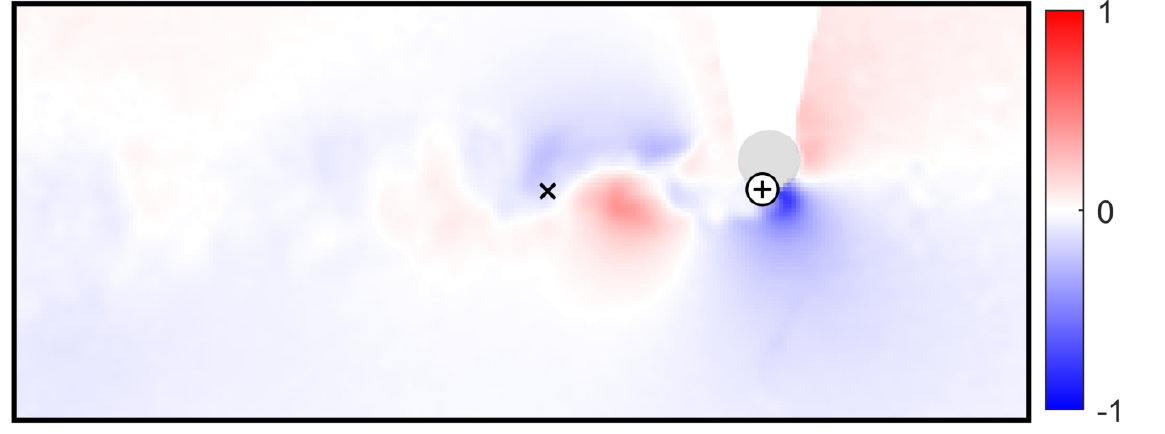}
		\caption{}
	\end{subfigure}
	\captionsetup{justification=raggedright,singlelinecheck=false}
	\caption{Phase averaged vorticity (top) and transverse velocity (bottom), $Re_0=900,$ $\sfrac{Re_q}{Re_0} = 0.35,$ $\sfrac{St_f}{St_0}=0.18$. (a) $\sfrac{t}{\tau_f}=0$; (b) $\sfrac{t}{\tau_f}=0.25$; (c) $\sfrac{t}{\tau_f}=0.50$; (d) $\sfrac{t}{\tau_f}=0.75$.}
	\label{fig:phaseAveragedComposite}
\end{figure}

Vorticity fields corresponding to the phase-averaged time-series are presented in \cref{fig:phaseAveragedComposite} for different phases in the cycle. The four snapshots of phase-averaged vorticity show the effect of forcing on shear-layer strength and spacing. As the cylinder moves downstream and the instantaneous freestream velocity seen by the cylinder decreases, the shear-layers weaken and move apart from each other. On the other hand, as the cylinder moves upstream, the instantaneous freestream velocity seen by the cylinder increases, and the shear-layers strengthen. This leads to the contraction and narrowing of the wake. As expected, phase-averaging reduces the observability of vortex shedding in the wake, the period for which is neither constant nor related to the forcing period through an integer number of cycles.

\begin{figure}
	\hspace{-0.75cm}
	\begin{subfigure}[b]{0.49\textwidth}
		\includegraphics[trim = 0px 51px 30px 95px,clip,scale = 0.5,center]{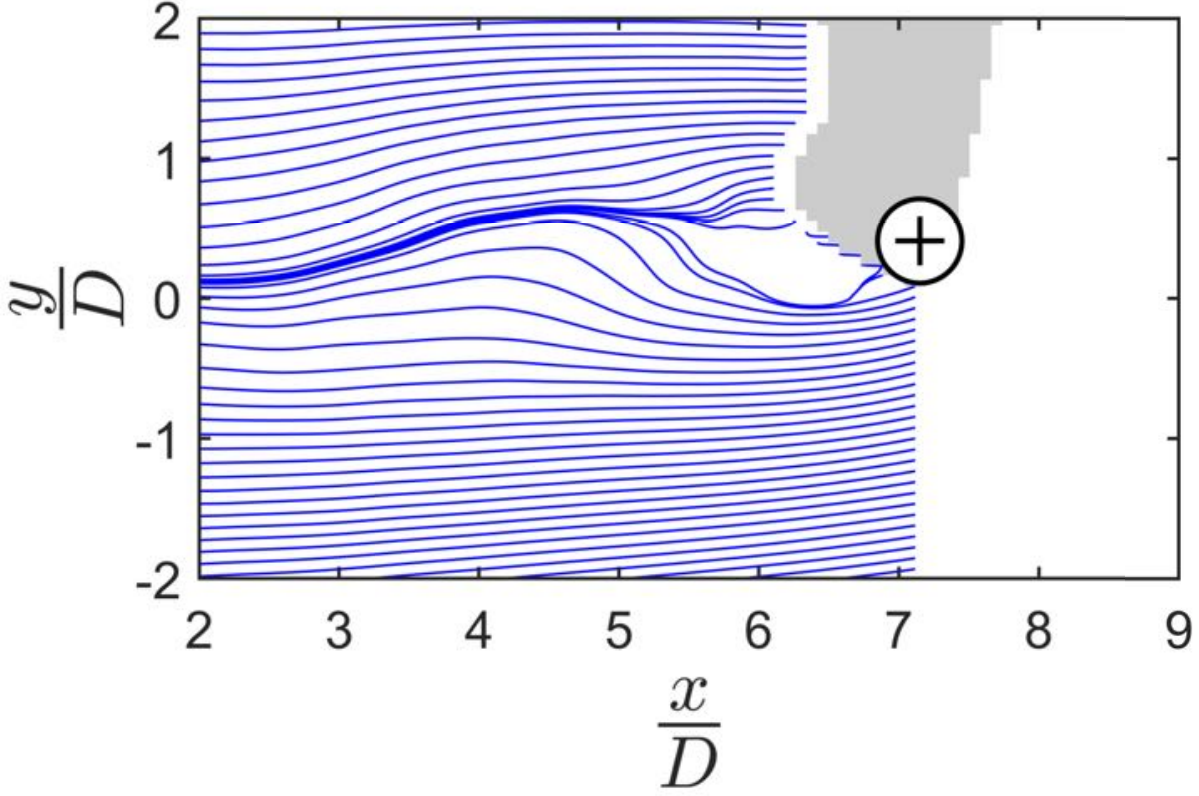}
		\label{fig:streamlines1}
	\end{subfigure}
	\begin{subfigure}[b]{0.49\textwidth}
		\includegraphics[trim =0px 51px 30px 95px,clip,scale = 0.5,left]{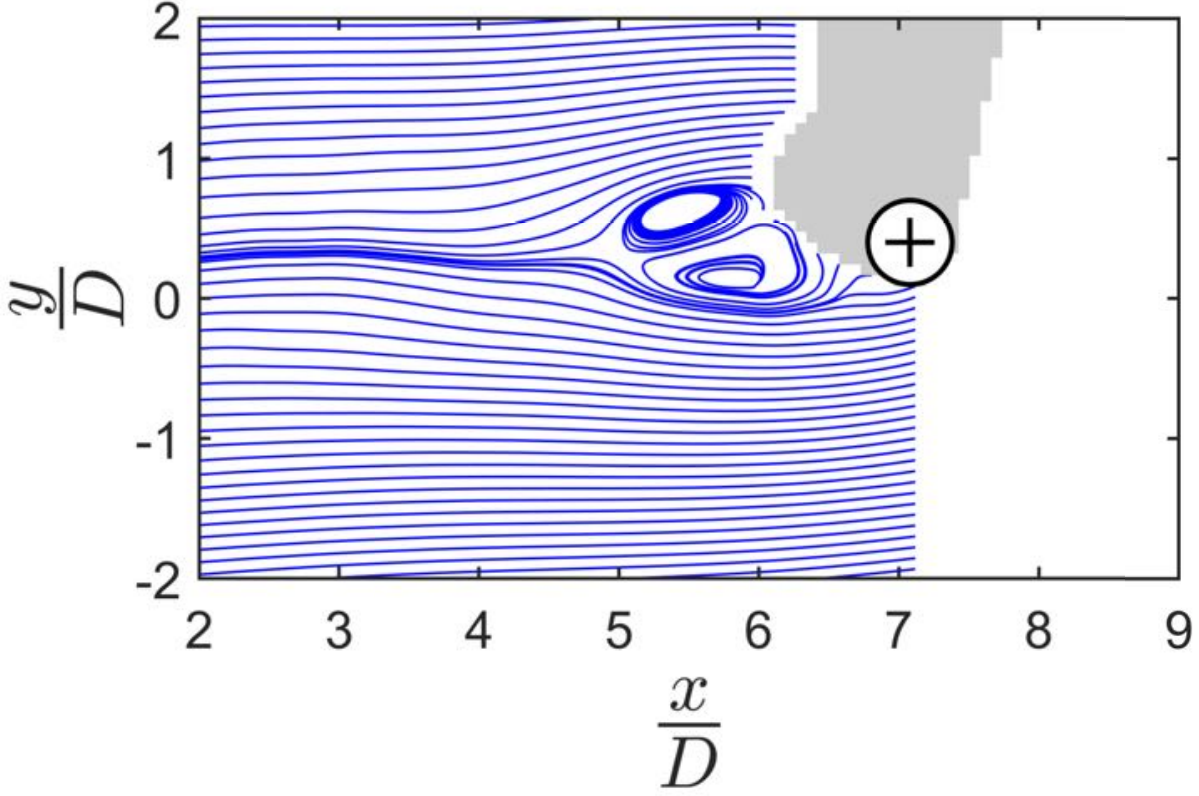}
		\label{fig:streamlines2}
	\end{subfigure}
	\captionsetup{justification=raggedright,singlelinecheck=false}
	\caption{Streamlines for phase averaged velocity field: before formation of starting vortex, $\frac{t}{\tau_f}=0.4$ (left); during formation of starting vortex, $\frac{t}{\tau_f}=0.5$ (right). $Re_0=900,$ $\sfrac{Re_q}{Re_0} = 0.35,$ $\sfrac{St_f}{St_0}=0.18$.}
	\label{fig:streamlines}
\end{figure}

\begin{figure}
	\begin{subfigure}[b]{0.24\textwidth}
		\includegraphics[trim =0 0 0 0, clip,scale=0.48]{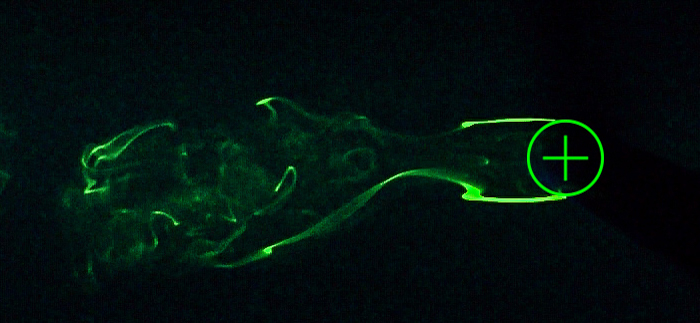}
		\caption{}
	\end{subfigure}
	\begin{subfigure}[b]{0.24\textwidth}
		\includegraphics[trim =0 0 0 0, clip,scale=0.48]{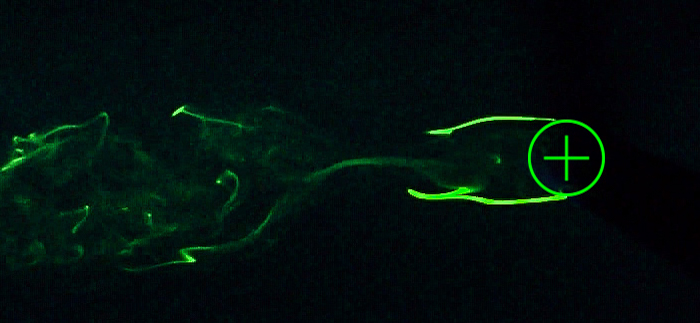}
		\caption{}
	\end{subfigure}
	\begin{subfigure}[b]{0.24\textwidth}
		\includegraphics[trim =0 0 0 0, clip,scale=0.48]{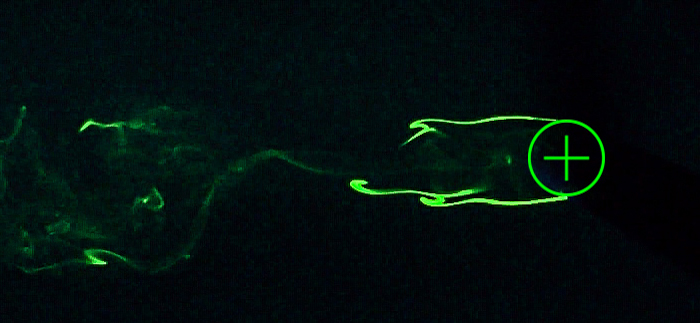}
		\caption{}
	\end{subfigure}
	\begin{subfigure}[b]{0.24\textwidth}
		\includegraphics[trim =0 0 0 0, clip,scale=0.48]{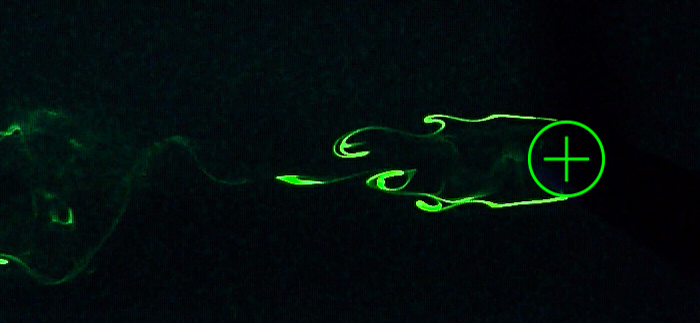}
		\caption{}
	\end{subfigure}
	\captionsetup{justification=raggedright,singlelinecheck=false}
	\caption{Dye flow visualization of starting vortex generation. (a) $\sfrac{t}{\tau_f}=0.50$; (b) $\sfrac{t}{\tau_f}=0.55$; (c) $\sfrac{t}{\tau_f}=0.60$; (d) $\sfrac{t}{\tau_f}=0.65$.}
	\label{fig:startingVortex}
\end{figure}

The streamlines of the phase-averaged flow, shown in \cref{fig:streamlines}, reveal a striking feature as the cylinder approaches the downstream turnaround point and changes direction, corresponding to $t = \frac{1}{2} \tau_f+k \tau_f$. Near the downstream turnaround point, initially smooth streamlines close in a symmetric manner about the wake centerline, before developing asymmetries as the forcing cycle advances and regular vortex shedding resumes. The development of closed streamlines alludes to the generation of two symmetric vortices downstream of the body with one on either side of the cylinder centerline, as seen clearly in the (instantaneous) dye-flow visualization of \cref{fig:startingVortex}. We refer to these as ``starting vortices" due to the near wake resemblance to that of an impulsively started cylinder.

\begin{figure}
	\captionsetup[subfigure]{oneside,margin={0.3cm,0cm}}
	\begin{subfigure}[b]{0.39\textwidth}
		\includegraphics[trim = 35px 15px 145px 190px,clip,scale = 0.87]{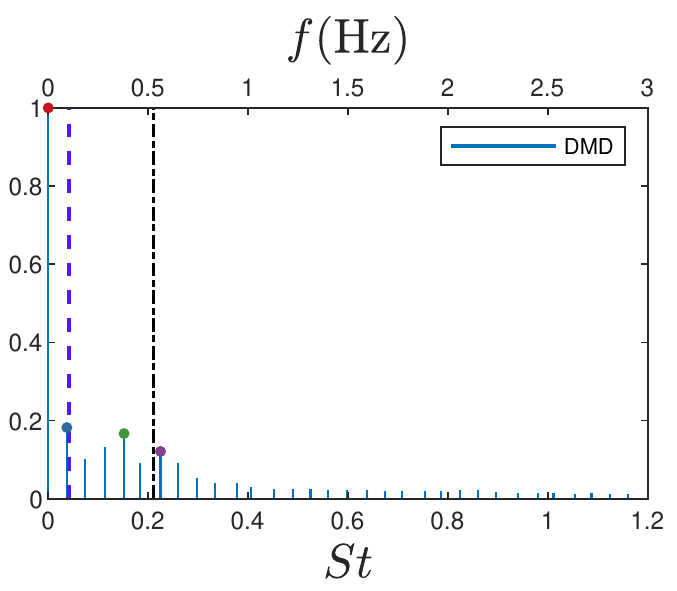}
		\caption{}
		\label{fig:phaseAveragedSpectrum}
	\end{subfigure}
	\captionsetup[subfigure]{oneside,margin={-0.1cm,0cm}}
	\begin{subfigure}[b]{0.55\textwidth}
		\includegraphics[trim = 20px 0 20px 0px,clip, scale = 1.35]{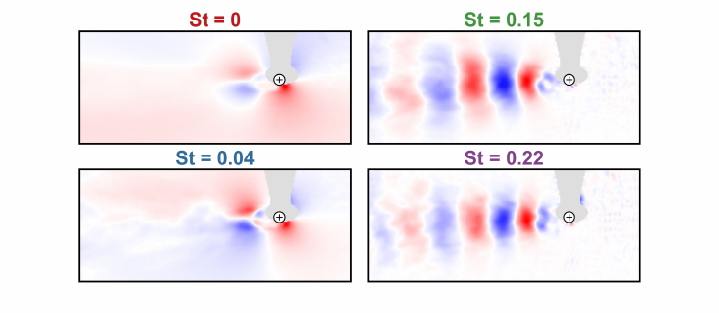}
		\vspace{0.4cm}
		\caption{}
		\label{fig:phaseAveragedModes}
	\end{subfigure}
	\captionsetup{justification=raggedright,singlelinecheck=false}
	\caption{KMD spectrum (a) and modes (b) for phase averaged time series, $Re_0=900,$ $\sfrac{Re_q}{Re_0} = 0.35,$ $\sfrac{St_f}{St_0}=0.18$.}
	\label{fig:phaseAveragedKMD}
\end{figure}

The Koopman Mode Decomposition of the phase averaged velocity fields is presented in \cref{fig:phaseAveragedKMD} (recall that the phase averaging implemented here leads to a lower frequency resolution in the KMD spectrum). Phase averaging leads to a reduction in activity in the frequency range for vortex shedding, a trend which should continue with increasing record length. The emergence of an anti-symmetric flow structure at the forcing frequency can be identified from \cref{fig:phaseAveragedModes}. This KMD mode will be referred to as the ``forcing mode". Footprints of the anti-symmetric forcing mode are also seen in the KMD presented above which were not phase-averaged (\cref{fig:forcedKMDModesFast}, \cref{fig:forcedKMDModesSlow}).  Furthermore, the phase averaged velocity field is directly related to the Koopman mode at the averaging frequency \cite{Mezic_2005}. Hence, the phase averaged transverse velocity (\cref{fig:phaseAveragedComposite}) also shows an imprint of the anti-symmetric flow structure representative of the forcing mode.

\subsubsection{Quasi-steady scaling}
\begin{figure}[t]
	\begin{subfigure}[b]{\textwidth}
		\includegraphics[trim = 170px 25px 0 180px, clip,scale = 0.63]{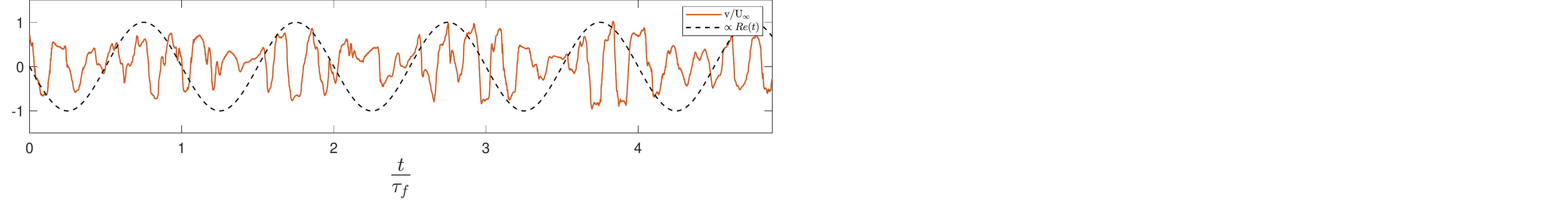}
	\end{subfigure}
	\begin{subfigure}[b]{\textwidth}
		\includegraphics[trim = 170px 25px 0 180px, clip,scale = 0.63]{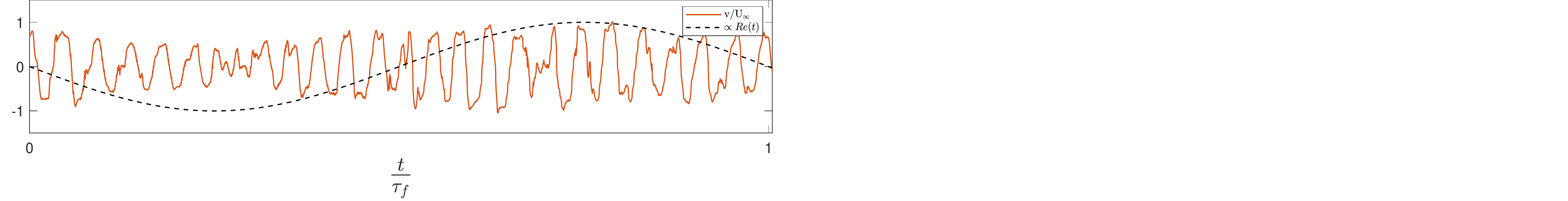}
		\label{fig:scaledTraceComparison_Slow}
	\end{subfigure}
	\captionsetup{justification=raggedright,singlelinecheck=false}
	\caption{Time-trace of transverse velocity, $\mathrm{v}/\mathrm{U}_\infty$, at interrogation point with quasi-steady time scaling. Top: $Re_0=900,$ $\sfrac{Re_q}{Re_0} = 0.35,$ $\sfrac{St_f}{St_0}=0.18$; bottom: $Re_0=900,$ $\sfrac{Re_q}{Re_0} = 0.35,$ $\sfrac{St_f}{St_0}=0.036$.}
	\label{fig:scaledTraceComparison}
\end{figure}

\begin{figure}[h]
	\captionsetup[subfigure]{oneside,margin={0.3cm,0cm}}
	\begin{subfigure}[b]{0.39\textwidth}
		\includegraphics[trim = 35px 15px 145px 190px,clip,scale = 0.87]{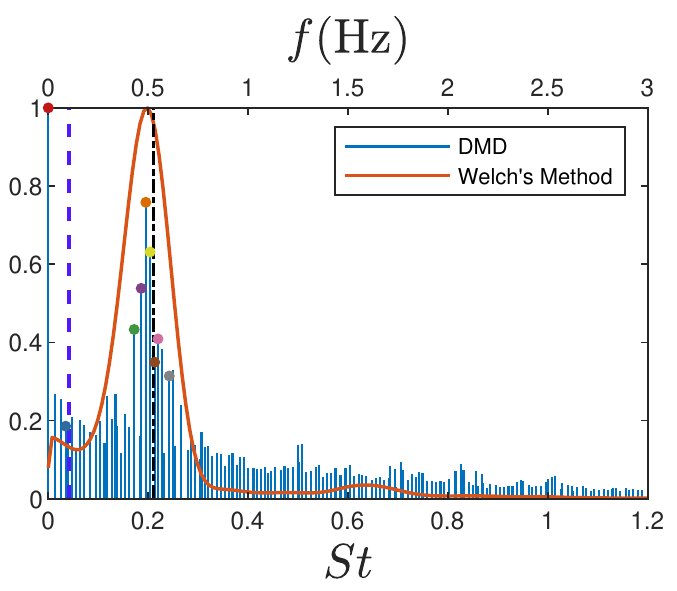}
		\caption{}
	\end{subfigure}
	\captionsetup[subfigure]{oneside,margin={-0.1cm,0cm}}
	\begin{subfigure}[b]{0.55\textwidth}
		\includegraphics[trim = 20px 0 20px 0px,clip, scale = 1.35]{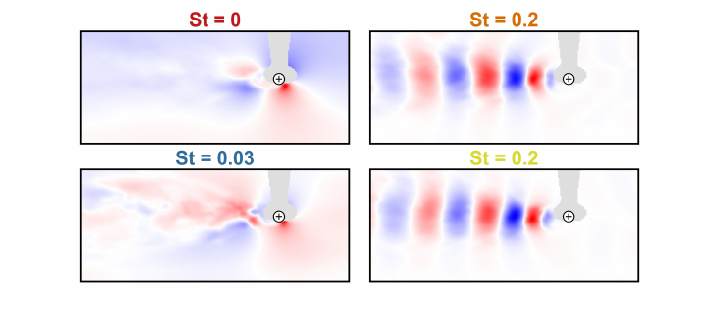}
		\vspace{0.4cm}
		\caption{}
	\end{subfigure}
	\captionsetup{justification=raggedright,singlelinecheck=false}
	\caption{KMD spectrum (a) and modes (b) for oscillating cylinder with quasi-steady time scaling, $Re_0=900,$ $\sfrac{Re_q}{Re_0} = 0.35,$ $\sfrac{St_f}{St_0}=0.18$.}
	\label{fig:scaledKMDFast}
\end{figure}

\begin{figure}[h]
	\captionsetup[subfigure]{oneside,margin={0.3cm,0cm}}
	\begin{subfigure}[b]{0.39\textwidth}
		\includegraphics[trim = 35px 15px 145px 190px,clip,scale = 0.783]{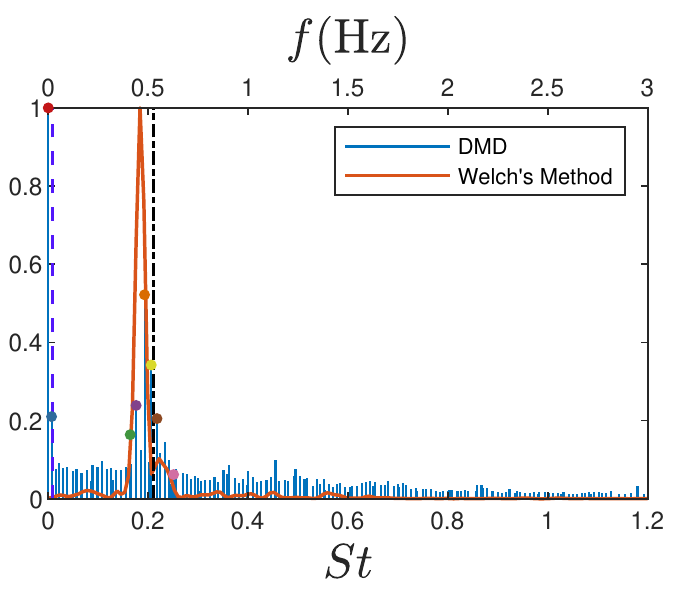}
		\caption{}
	\end{subfigure}
	\captionsetup[subfigure]{oneside,margin={-0.1cm,0cm}}
	\begin{subfigure}[b]{0.55\textwidth}
		\includegraphics[trim = 0px 0 0px 0px,clip, scale = 1.215]{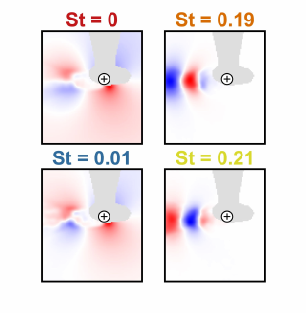}
		\vspace{0.4cm}
		\caption{}
	\end{subfigure}
	\captionsetup{justification=raggedright,singlelinecheck=false}
	\caption{KMD spectrum (a) and modes (b) for oscillating cylinder with quasi-steady time scaling, $Re_0=900,$ $\sfrac{Re_q}{Re_0} = 0.35,$ $\sfrac{St_f}{St_0}=0.036$.}
	\label{fig:scaledKMDSlow}
\end{figure}

We now consider the flow field and KMD under the quasi-steady time scaling introduced in \cref{section:timescaling}. \Cref{fig:scaledTraceComparison} shows the time traces of \cref{fig:timeScaling}, which exhibited strong frequency modulation, under time scaling, i.e. with respect to  $\tilde{t}$. For both forcing regimes an approximately constant shedding frequency can be observed, which is particularly evident for $St_f\ll St_0$. This observation is confirmed by the PSDs in \cref{fig:scaledKMDFast} and \cref{fig:scaledKMDSlow}, which show a single peak and significant narrowing from the two peak spectra of \cref{fig:forcedKMDFast} and \cref{fig:forcedKMDSpectrumSlow}, respectively.

KMD of the scaled data, presented in \cref{fig:scaledKMDFast,fig:scaledKMDSlow}, also shows that the two peaks present in the spectra of the unscaled cases are no longer present. Furthermore, KMD extracts fewer shedding modes which are located in a smaller band of frequencies when compared to the unscaled KMD. In addition to the collapse of shedding modes onto a single peak, the prominence of the peak is also increased relative to the unscaled case. While the quasi-steady scaling seems to be relatively successful in identifying a simpler structure to the vortex shedding, it cannot treat the development of the starting vortices.

\section{Discussion}\label{section:Discussion}

The previous section has identified the development of numerous unsteady phenomena in the streamwise-oscillating cylinder wake, including frequency modulation, amplitude modulation, and the generation of starting vortices. The complexity of this flow can be reduced by quasi-steady time scaling, the success of which hinges on two factors. First, the characteristic (shedding) frequency of the stationary cylinder must be primarily governed by the Reynolds number. Second, the separation of scales between forcing and shedding frequency facilitates the development of quasi-steady shedding during certain portions of the forcing cycle. These two factors connect the forcing trajectory to the frequency modulation observed.

\begin{figure}[t]
	\includegraphics[trim = 16px 14px 0 220px,clip]{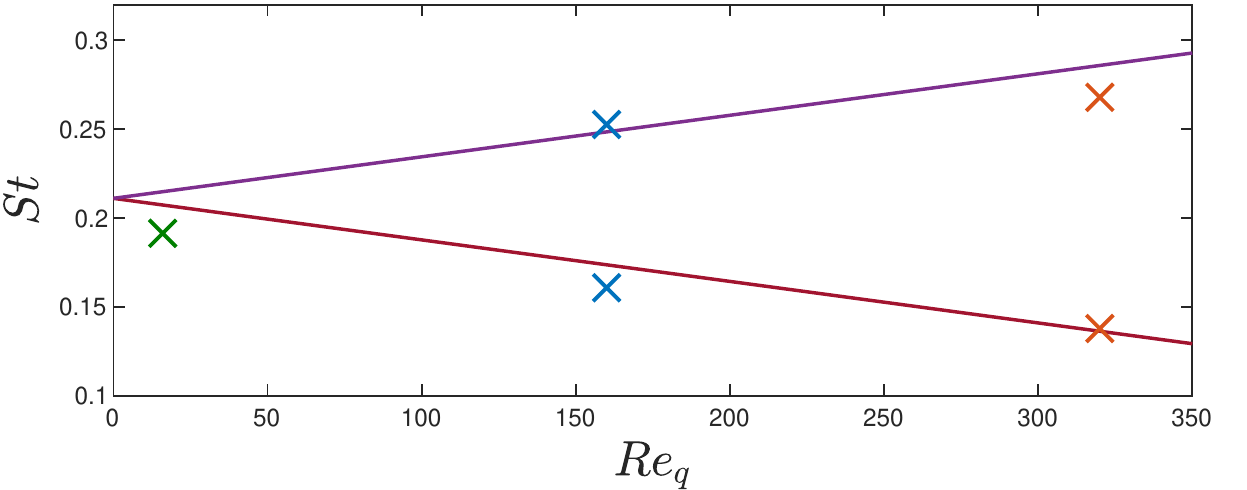}
	\captionsetup{justification=raggedright,singlelinecheck=false}
	\caption{Comparison of predicted and observed dominant frequencies, $Re_0=900,$ $\sfrac{St_f}{St_0}=0.18$. The solid lines represent the predicted dominant frequencies while the $\mathbf{\times}$ symbols represent the actual dominant frequencies observed in experiments.}
	\label{fig:prediction}
\end{figure}

As presented in \cref{section:analysis}, the system is predicted to show strong quasi-steady behavior when $\Omega$ is large. Using \cref{eqn:frequency} the shedding frequency corresponding to each of the two peaks in $\Omega$ can be predicted. In \cref{fig:prediction} the observed spectral peaks for the interrogation point are compared with the (predicted) dominant Strouhal number corresponding to each peak in $\Omega$. The peaks in the interrogation point spectra are in good agreement with the prediction indicating that the two peak nature of the unscaled PSDs can be identified with periods of large $\Omega$ and that the observed dominant frequencies arise due to quasi-steady behavior.

\begin{figure}[b]
	\begin{subfigure}[b]{0.49\textwidth}
			\includegraphics[trim = 35px 15px 145px 217px,clip]{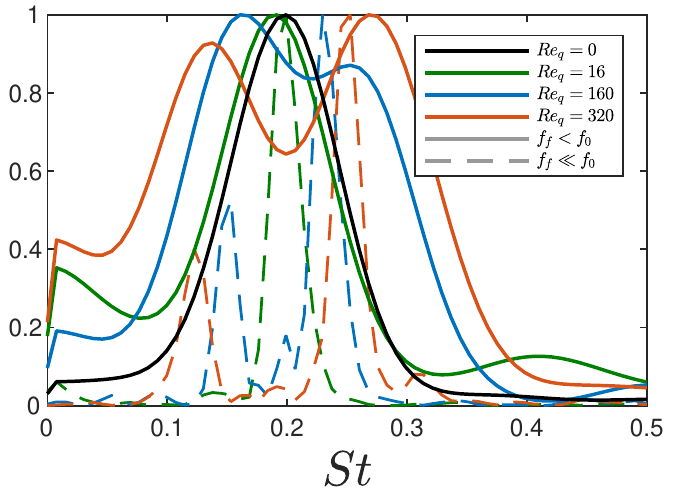}
	\end{subfigure}
	\begin{subfigure}[b]{0.49\textwidth}
			\includegraphics[trim = 35px 15px 145px 217px,clip]{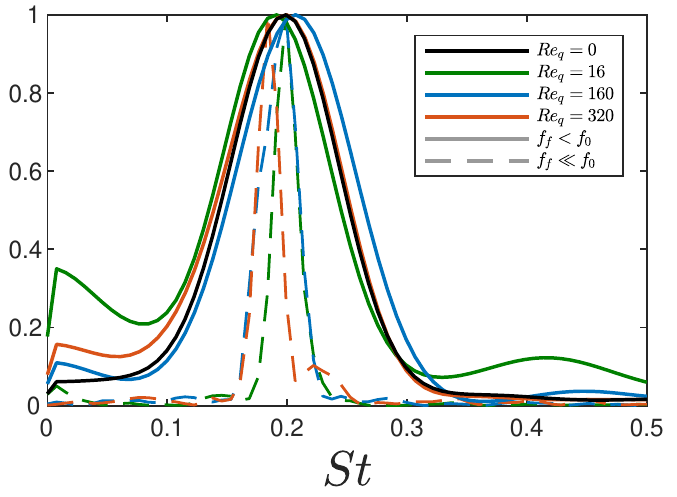}
	\end{subfigure}
	\captionsetup{justification=raggedright,singlelinecheck=false}
	\caption{Comparison of interrogation point power spectra before (left) and after (right) application of quasi-steady time scaling.}
	\label{fig:collapse}
\end{figure}

The spectral peaks from a range of forcing amplitudes, $Re_q$, exhibit quasi-steady behavior, as shown in \cref{fig:collapse}. The peaks are narrower for $St_f\ll St_0$; conceptually, the wake has more time to settle onto frequencies corresponding to the instantaneous Reynolds number, i.e. is more quasi-steady, for this case.

Although the dominant frequencies predicted in \cref{fig:prediction} are in good agreement with observed spectra, the assumptions were rather crude, thus a few factors may be underlying any discrepancies. Firstly, the spectra were calculated using Welch's method, which depends heavily on window quantity and size. Varying these parameters can lead to different estimates of PSD. Thus, these parameters may cause the interrogation point spectra to deviate from the prediction. Although $St_0$ is approximately constant for the range of Reynolds numbers encountered, it can vary by up to 3\% during the forcing cycle, a deviation which may percolate through the analysis. It should also be noted time scaling is applied uniformly to each snapshot in the time series. That is, the scaling procedure does not account for the fact that it takes a finite amount of time for vortical structures to convect downstream. Even still, the quasi-steady time scaling gives reasonable collapse of the two peaks in the measured spectra onto a single spectral peak.
Finally, it should be mentioned that the scaling procedure does not account for the vortex interactions, shear-layer dynamics, and resulting amplitude modulation discussed above. These factors may further cause deviations from the prediction. For example, the modulation of shear-layer strength is such that starting vortices may develop at times close to peaks in $\Omega$, where shedding at the dominant frequencies is expected.

Due to the continuous change in shedding frequency along a forcing trajectory, KMD extracts additional shedding modes at multiple frequencies, in addition to those at the two dominant frequencies. This means that the dynamics of the unscaled system are not restricted to a small number of shedding modes and frequencies, even though the shape and structure of the shedding modes are quite similar.

Although the forcing trajectory leads to some amplitude modulation in the wake the mechanism governing this behavior is not identical to that responsible for frequency modulation. As presented above, the development of symmetric starting vortices and amplitude modulation are both phase-locked to the forcing. Furthermore, phase-averaged vorticity fields show that the modulation of shear-layer strength is also phase-locked to the forcing. Consideration of the relative freestream velocity seen by the cylinder helps elucidate the cause of this phenomena. As the cylinder moves downstream the relative freestream velocity seen by the cylinder decreases and so does the strength of the shear-layer. The cylinder follows a sinusoidal forcing trajectory so as it starts to approach the downstream turnaround point, it slows down, causing an increase in the relative freestream velocity seen by the cylinder. Consequently the shear-layer strength begins to increase. When the shear-layer becomes sufficiently strong, starting vortices are generated. As the starting vortices develop, asymmetric shedding ceases, and the wake is relatively quiescent (i.e., oscillatory behavior and transverse velocity are suppressed). Thus, the spatio-temporal variation of the shedding type and strength leads to amplitude modulation in the wake.

The range of forcing frequencies considered in this study do not overlap with those required for the canonical lock-on phenomena.  The generation of starting vortices at the downstream turnaround point, however, is synchronized with the forcing. In contrast to the classical definition of lock-on in which the vortex shedding frequency is proportional to the forcing frequency, the generation of starting vortices can be thought of as a ``pseudo" or ``partial" lock-on phenomenon. Specifically, the generation frequency of starting vortices is proportional to the forcing frequency even though the vortex shedding frequency is not.

\section{Conclusion}\
\label{section:Conclusion}
The flow around streamwise-oscillating cylinders at moderate Reynolds number was studied. Forcing frequencies one and two orders of magnitude less than the stationary shedding frequency were considered, well below the frequency ratios corresponding to lock-on. Various amplitude ratios were also studied, all satisfying the requirement that the instantaneous Reynolds number remain above the critical value. Experiments were carried out in a closed loop water tunnel in order to assess the effect of streamwise forcing on the wake. Particle Image Velocimetry was used to obtain snapshots of velocity while fluorescent dye was used to illuminate flow structures that developed due to streamwise forcing. Experiments showed that streamwise forcing led to the development of both frequency and amplitude modulation in the wake. Additionally, a range of vortex interactions were observed including the development of symmetric starting vortices.

Koopman analysis was used to extract spatio-temporally significant flow structures from experimental velocity fields. This revealed that vortex shedding lead to the generation of shedding modes at the corresponding shedding frequency. Furthermore, it was seen that as the amplitude of oscillation increased, the number of shedding modes extracted using Koopman analysis followed suit. Global spectra from Koopman analysis revealed a spread in the spectral distribution of shedding modes. Additionally, interrogation point spectra revealed the development of two spectral peaks, indicating the presence of two dominant frequencies in the streamwise-oscillating cylinder's wake.

Phase-averaging was used to show that shear-layer strength was phase-locked to the forcing. It was shown that the modulation of shear-layer strength lead to the development of symmetric starting vortices at the downstream turnaround point. Consequently, oscillatory behavior in the wake was suppressed, leading to amplitude modulation. Koopman analysis of the phase-averaged flow-fields resulted in the extraction of a forcing mode corresponding to the modulation of shear-layer strength.

In addition to experiments, analysis was used to show that the forcing frequency allowed for the development of quasi-steady dynamics at various points in the forcing trajectory. A parameter, the quasi-steadiness, was developed in order to predict when quasi-steady dynamics would be present and was in good agreement with experiments. Furthermore, it was shown that the shedding frequency of the system could be predicted for portions of the forcing cycle exhibiting quasi-steady behavior. Using the interplay between unsteady and quasi-steady dynamics along the forcing trajectory, a transformation was developed to scale time and normalize the characteristic frequency of the system. That is, the time scaling was used to transform the system's dynamics from unsteady to quasi-steady. Time scaling applied to experimental velocity fields showed that both the Koopman and interrogation point spectra collapsed onto a single spectral peak. Furthermore, fewer shedding modes were extracted, indicating that vortex shedding was occurring in smaller band of frequencies. 

It was shown that constant sampling relative to the lab time, $t$, resulted in and irregular sampling rate relative to scaled time, $\tilde{t}$. Going forward, experiments can be performed with constant sampling relative to the scaled time in order to bypass the need for time scaling as a step in post-processing. It is worth noting that the analysis presented in this work is not limited to fluid systems, but rather it is applicable to any periodic system with a characteristic frequency determined by a set of governing parameters. Many systems of engineering and scientific interest display periodic behavior. Of these, multiple have well established connections between governing parameters and frequency. These systems may indeed be amenable to the time scaling and analysis presented here.

\begin{acknowledgments}
This work was supported by U.S. Army Research Office grant \# W911NF-17-1-0306. The authors would also like to thank Alexandra Techet, Shai Revzen, and Morgan Hooper for insightful discussions and input.
\end{acknowledgments}

\nocite{Arbabi_2017,Williamson_1988, Williamson_2004,Leontini_2011,Leontini_2013,Glaz_2017}


\begin{thebibliography}{17}%
\makeatletter
\providecommand \@ifxundefined [1]{%
 \@ifx{#1\undefined}
}%
\providecommand \@ifnum [1]{%
 \ifnum #1\expandafter \@firstoftwo
 \else \expandafter \@secondoftwo
 \fi
}%
\providecommand \@ifx [1]{%
 \ifx #1\expandafter \@firstoftwo
 \else \expandafter \@secondoftwo
 \fi
}%
\providecommand \natexlab [1]{#1}%
\providecommand \enquote  [1]{``#1''}%
\providecommand \bibnamefont  [1]{#1}%
\providecommand \bibfnamefont [1]{#1}%
\providecommand \citenamefont [1]{#1}%
\providecommand \href@noop [0]{\@secondoftwo}%
\providecommand \href [0]{\begingroup \@sanitize@url \@href}%
\providecommand \@href[1]{\@@startlink{#1}\@@href}%
\providecommand \@@href[1]{\endgroup#1\@@endlink}%
\providecommand \@sanitize@url [0]{\catcode `\\12\catcode `\$12\catcode
  `\&12\catcode `\#12\catcode `\^12\catcode `\_12\catcode `\%12\relax}%
\providecommand \@@startlink[1]{}%
\providecommand \@@endlink[0]{}%
\providecommand \url  [0]{\begingroup\@sanitize@url \@url }%
\providecommand \@url [1]{\endgroup\@href {#1}{\urlprefix }}%
\providecommand \urlprefix  [0]{URL }%
\providecommand \Eprint [0]{\href }%
\providecommand \doibase [0]{https://doi.org/}%
\providecommand \selectlanguage [0]{\@gobble}%
\providecommand \bibinfo  [0]{\@secondoftwo}%
\providecommand \bibfield  [0]{\@secondoftwo}%
\providecommand \translation [1]{[#1]}%
\providecommand \BibitemOpen [0]{}%
\providecommand \bibitemStop [0]{}%
\providecommand \bibitemNoStop [0]{.\EOS\space}%
\providecommand \EOS [0]{\spacefactor3000\relax}%
\providecommand \BibitemShut  [1]{\csname bibitem#1\endcsname}%
\let\auto@bib@innerbib\@empty
\bibitem [{\citenamefont {C.Barbi}\ \emph {et~al.}(1986)\citenamefont
  {C.Barbi}, \citenamefont {Favier}, \citenamefont {Maresca},\ and\
  \citenamefont {Telionis}}]{Barbi_1986}%
  \BibitemOpen
  \bibfield  {author} {\bibinfo {author} {\bibnamefont {C.Barbi}}, \bibinfo
  {author} {\bibfnamefont {D.}~\bibnamefont {Favier}}, \bibinfo {author}
  {\bibfnamefont {C.}~\bibnamefont {Maresca}},\ and\ \bibinfo {author}
  {\bibfnamefont {D.}~\bibnamefont {Telionis}},\ }\bibfield  {title} {\bibinfo
  {title} {Vortex shedding and lock-on of a circular cylinder in oscillatory
  flow},\ }\href@noop {} {\bibfield  {journal} {\bibinfo  {journal} {Journal of
  Fluid Mechanics}\ }\textbf {\bibinfo {volume} {170}},\ \bibinfo {pages} {527}
  (\bibinfo {year} {1986})}\BibitemShut {NoStop}%
\bibitem [{\citenamefont {Griffin}\ and\ \citenamefont
  {Ramberg}(1976)}]{Griffin_1976}%
  \BibitemOpen
  \bibfield  {author} {\bibinfo {author} {\bibfnamefont {O.}~\bibnamefont
  {Griffin}}\ and\ \bibinfo {author} {\bibfnamefont {S.}~\bibnamefont
  {Ramberg}},\ }\bibfield  {title} {\bibinfo {title} {Vortex shedding from a
  cylinder vibrating in line with an incident uniform flow},\ }\href@noop {}
  {\bibfield  {journal} {\bibinfo  {journal} {Journal of Fluid Mechanics}\
  }\textbf {\bibinfo {volume} {75}},\ \bibinfo {pages} {257} (\bibinfo {year}
  {1976})}\BibitemShut {NoStop}%
\bibitem [{\citenamefont {Leontini}\ \emph {et~al.}(2011)\citenamefont
  {Leontini}, \citenamefont {Jacono},\ and\ \citenamefont
  {Thompson}}]{Leontini_2011}%
  \BibitemOpen
  \bibfield  {author} {\bibinfo {author} {\bibfnamefont {J.}~\bibnamefont
  {Leontini}}, \bibinfo {author} {\bibfnamefont {D.}~\bibnamefont {Jacono}},\
  and\ \bibinfo {author} {\bibfnamefont {M.}~\bibnamefont {Thompson}},\
  }\bibfield  {title} {\bibinfo {title} {A numerical study of an inline
  oscillating cylinder in a freestream},\ }\href@noop {} {\bibfield  {journal}
  {\bibinfo  {journal} {Journal of Fluid Mechanics}\ }\textbf {\bibinfo
  {volume} {688}},\ \bibinfo {pages} {551} (\bibinfo {year}
  {2011})}\BibitemShut {NoStop}%
\bibitem [{\citenamefont {Leontini}\ \emph {et~al.}(2013)\citenamefont
  {Leontini}, \citenamefont {Jacono},\ and\ \citenamefont
  {Thompson}}]{Leontini_2013}%
  \BibitemOpen
  \bibfield  {author} {\bibinfo {author} {\bibfnamefont {J.}~\bibnamefont
  {Leontini}}, \bibinfo {author} {\bibfnamefont {D.}~\bibnamefont {Jacono}},\
  and\ \bibinfo {author} {\bibfnamefont {M.}~\bibnamefont {Thompson}},\
  }\bibfield  {title} {\bibinfo {title} {Wake states and frequency selection of
  a streamwise oscillating cylinder},\ }\href@noop {} {\bibfield  {journal}
  {\bibinfo  {journal} {Journal of Fluid Mechanics}\ }\textbf {\bibinfo
  {volume} {730}},\ \bibinfo {pages} {162} (\bibinfo {year}
  {2013})}\BibitemShut {NoStop}%
\bibitem [{\citenamefont {Rowley}\ and\ \citenamefont
  {Dawson}(2017)}]{Rowley_2017}%
  \BibitemOpen
  \bibfield  {author} {\bibinfo {author} {\bibfnamefont {C.}~\bibnamefont
  {Rowley}}\ and\ \bibinfo {author} {\bibfnamefont {S.}~\bibnamefont
  {Dawson}},\ }\bibfield  {title} {\bibinfo {title} {Model reduction for flow
  analysis and control},\ }\href@noop {} {\bibfield  {journal} {\bibinfo
  {journal} {Annual Review of Fluid Mechanics}\ }\textbf {\bibinfo {volume}
  {49}},\ \bibinfo {pages} {387} (\bibinfo {year} {2017})}\BibitemShut
  {NoStop}%
\bibitem [{\citenamefont {Taira}\ \emph {et~al.}(2017)\citenamefont {Taira},
  \citenamefont {Brunton}, \citenamefont {Dawson}, \citenamefont {Rowley},
  \citenamefont {Colonius}, \citenamefont {McKeon}, \citenamefont {Schmidt},
  \citenamefont {Gordeyev}, \citenamefont {Theofilis},\ and\ \citenamefont
  {Ukeiley}}]{Taira_2017}%
  \BibitemOpen
  \bibfield  {author} {\bibinfo {author} {\bibfnamefont {K.}~\bibnamefont
  {Taira}}, \bibinfo {author} {\bibfnamefont {S.}~\bibnamefont {Brunton}},
  \bibinfo {author} {\bibfnamefont {S.}~\bibnamefont {Dawson}}, \bibinfo
  {author} {\bibfnamefont {C.}~\bibnamefont {Rowley}}, \bibinfo {author}
  {\bibfnamefont {T.}~\bibnamefont {Colonius}}, \bibinfo {author}
  {\bibfnamefont {B.}~\bibnamefont {McKeon}}, \bibinfo {author} {\bibfnamefont
  {O.}~\bibnamefont {Schmidt}}, \bibinfo {author} {\bibfnamefont
  {S.}~\bibnamefont {Gordeyev}}, \bibinfo {author} {\bibfnamefont
  {V.}~\bibnamefont {Theofilis}},\ and\ \bibinfo {author} {\bibfnamefont
  {L.}~\bibnamefont {Ukeiley}},\ }\bibfield  {title} {\bibinfo {title} {Modal
  analysis of fluid flows: An overview},\ }\href@noop {} {\bibfield  {journal}
  {\bibinfo  {journal} {AIAA Journal}\ }\textbf {\bibinfo {volume} {55}},\
  \bibinfo {pages} {4013} (\bibinfo {year} {2017})}\BibitemShut {NoStop}%
\bibitem [{\citenamefont {Rowley}\ \emph {et~al.}(2009)\citenamefont {Rowley},
  \citenamefont {Mezi\'{c}}, \citenamefont {Bagheri}, \citenamefont
  {Schlatter},\ and\ \citenamefont {s~Henningson}}]{Rowley_2009}%
  \BibitemOpen
  \bibfield  {author} {\bibinfo {author} {\bibfnamefont {C.~W.}\ \bibnamefont
  {Rowley}}, \bibinfo {author} {\bibfnamefont {I.}~\bibnamefont {Mezi\'{c}}},
  \bibinfo {author} {\bibfnamefont {S.}~\bibnamefont {Bagheri}}, \bibinfo
  {author} {\bibfnamefont {P.}~\bibnamefont {Schlatter}},\ and\ \bibinfo
  {author} {\bibfnamefont {D.}~\bibnamefont {s~Henningson}},\ }\bibfield
  {title} {\bibinfo {title} {Spectral analysis of nonlinear flows},\ }\href
  {https://doi.org/https://doi.org/10.1017/S0022112009992059} {\bibfield
  {journal} {\bibinfo  {journal} {Journal of Fluid Mechanics}\ }\textbf
  {\bibinfo {volume} {641}},\ \bibinfo {pages} {115} (\bibinfo {year}
  {2009})}\BibitemShut {NoStop}%
\bibitem [{\citenamefont {Tu}\ \emph {et~al.}(2011)\citenamefont {Tu},
  \citenamefont {Rowley}, \citenamefont {Aram},\ and\ \citenamefont
  {Mittal}}]{Tu_2011}%
  \BibitemOpen
  \bibfield  {author} {\bibinfo {author} {\bibfnamefont {J.}~\bibnamefont
  {Tu}}, \bibinfo {author} {\bibfnamefont {C.}~\bibnamefont {Rowley}}, \bibinfo
  {author} {\bibfnamefont {E.}~\bibnamefont {Aram}},\ and\ \bibinfo {author}
  {\bibfnamefont {R.}~\bibnamefont {Mittal}},\ }\bibfield  {title} {\bibinfo
  {title} {Koopman spectral analysis of separated flow over a finite-thickness
  flat plate with elliptical leading edge},\ }\href@noop {} {\bibfield
  {journal} {\bibinfo  {journal} {49th AIAA Aerospace Sciences Meeting
  including the New Horizons Forum and Aerospace Exposition}\ ,\ \bibinfo
  {pages} {1}} (\bibinfo {year} {2011})}\BibitemShut {NoStop}%
\bibitem [{\citenamefont {Haykin}(2001)}]{Haykin_2001}%
  \BibitemOpen
  \bibfield  {author} {\bibinfo {author} {\bibfnamefont {S.}~\bibnamefont
  {Haykin}},\ }\bibinfo {title} {Communication {S}ystems}\ (\bibinfo
  {publisher} {John {W}iley \& {S}ons},\ \bibinfo {address} {New York, New
  York},\ \bibinfo {year} {2001})\ pp.\ \bibinfo {pages} {107--111}\BibitemShut
  {NoStop}%
\bibitem [{\citenamefont {Fey}\ \emph {et~al.}(1998)\citenamefont {Fey},
  \citenamefont {K{\"o}nig},\ and\ \citenamefont {Eckelmann}}]{Fey_1998}%
  \BibitemOpen
  \bibfield  {author} {\bibinfo {author} {\bibfnamefont {U.}~\bibnamefont
  {Fey}}, \bibinfo {author} {\bibfnamefont {M.}~\bibnamefont {K{\"o}nig}},\
  and\ \bibinfo {author} {\bibfnamefont {H.}~\bibnamefont {Eckelmann}},\
  }\bibfield  {title} {\bibinfo {title} {A new {S}trouhal-{R}eynolds number
  relationship for the circular cylinder in the range
  {$47<\mathrm{Re}<2\times10^5$}},\ }\href
  {https://doi.org/http://dx.doi.org/10.1063/1.869675} {\bibfield  {journal}
  {\bibinfo  {journal} {Physics of Fluids}\ }\textbf {\bibinfo {volume} {10}},\
  \bibinfo {pages} {1547} (\bibinfo {year} {1998})}\BibitemShut {NoStop}%
\bibitem [{\citenamefont {Schmid}(2010)}]{Schmid_2010}%
  \BibitemOpen
  \bibfield  {author} {\bibinfo {author} {\bibfnamefont {P.~J.}\ \bibnamefont
  {Schmid}},\ }\bibfield  {title} {\bibinfo {title} {Dynamic mode decomposition
  of numerical and experimental data},\ }\href
  {https://doi.org/https://doi.org/10.1017/S0022112010001217} {\bibfield
  {journal} {\bibinfo  {journal} {Journal of Fluid Mechanics}\ }\textbf
  {\bibinfo {volume} {656}},\ \bibinfo {pages} {5} (\bibinfo {year}
  {2010})}\BibitemShut {NoStop}%
\bibitem [{\citenamefont {Tu}\ \emph {et~al.}(2014)\citenamefont {Tu},
  \citenamefont {Rowley}, \citenamefont {Luchtenburg}, \citenamefont
  {Brunton},\ and\ \citenamefont {Kutz}}]{Tu_2014}%
  \BibitemOpen
  \bibfield  {author} {\bibinfo {author} {\bibfnamefont {J.~H.}\ \bibnamefont
  {Tu}}, \bibinfo {author} {\bibfnamefont {C.~W.}\ \bibnamefont {Rowley}},
  \bibinfo {author} {\bibfnamefont {D.~M.}\ \bibnamefont {Luchtenburg}},
  \bibinfo {author} {\bibfnamefont {S.~L.}\ \bibnamefont {Brunton}},\ and\
  \bibinfo {author} {\bibfnamefont {J.~N.}\ \bibnamefont {Kutz}},\ }\bibfield
  {title} {\bibinfo {title} {On dynamic mode decomposition: Theory and
  applications},\ }\href
  {https://doi.org/http://dx.doi.org/10.3934/jcd.2014.1.391} {\bibfield
  {journal} {\bibinfo  {journal} {Journal of Computational Dynamics}\ }\textbf
  {\bibinfo {volume} {1}},\ \bibinfo {pages} {391} (\bibinfo {year}
  {2014})}\BibitemShut {NoStop}%
\bibitem [{\citenamefont {Mezi\'{c}}(2005)}]{Mezic_2005}%
  \BibitemOpen
  \bibfield  {author} {\bibinfo {author} {\bibfnamefont {I.}~\bibnamefont
  {Mezi\'{c}}},\ }\bibfield  {title} {\bibinfo {title} {{S}pectral properties
  of dynamical systems, model reduction, and decomposition},\ }\href@noop {}
  {\bibfield  {journal} {\bibinfo  {journal} {Nonlinear Dynamics}\ }\textbf
  {\bibinfo {volume} {41}},\ \bibinfo {pages} {309} (\bibinfo {year}
  {2005})}\BibitemShut {NoStop}%
\bibitem [{\citenamefont {Arbabi}\ and\ \citenamefont
  {Mezi\'{c}}(2017)}]{Arbabi_2017}%
  \BibitemOpen
  \bibfield  {author} {\bibinfo {author} {\bibfnamefont {H.}~\bibnamefont
  {Arbabi}}\ and\ \bibinfo {author} {\bibfnamefont {I.}~\bibnamefont
  {Mezi\'{c}}},\ }\bibfield  {title} {\bibinfo {title} {Study of dynamics in
  post-transient flows using {K}oopman mode decomposition},\ }\href@noop {}
  {\bibfield  {journal} {\bibinfo  {journal} {Physical Review Fluids}\ }\textbf
  {\bibinfo {volume} {2}},\ \bibinfo {pages} {124402} (\bibinfo {year}
  {2017})}\BibitemShut {NoStop}%
\bibitem [{\citenamefont {Williamson}\ and\ \citenamefont
  {Roshko}(1988)}]{Williamson_1988}%
  \BibitemOpen
  \bibfield  {author} {\bibinfo {author} {\bibfnamefont {C.}~\bibnamefont
  {Williamson}}\ and\ \bibinfo {author} {\bibfnamefont {A.}~\bibnamefont
  {Roshko}},\ }\bibfield  {title} {\bibinfo {title} {Vortex formation in the
  wake of an oscillating cylinder},\ }\href
  {https://doi.org/https://doi.org/10.1016/S0889-9746(88)90058} {\bibfield
  {journal} {\bibinfo  {journal} {Journal of Fluids and Structures}\ }\textbf
  {\bibinfo {volume} {2}},\ \bibinfo {pages} {351} (\bibinfo {year}
  {1988})}\BibitemShut {NoStop}%
\bibitem [{\citenamefont {Williamson}\ and\ \citenamefont
  {Govardhan}(2004)}]{Williamson_2004}%
  \BibitemOpen
  \bibfield  {author} {\bibinfo {author} {\bibfnamefont {C.}~\bibnamefont
  {Williamson}}\ and\ \bibinfo {author} {\bibfnamefont {R.}~\bibnamefont
  {Govardhan}},\ }\bibfield  {title} {\bibinfo {title} {Vortex-induced
  vibrations},\ }\href
  {https://doi.org/https://doi.org/10.1146/annurev.fluid.36.050802.122128}
  {\bibfield  {journal} {\bibinfo  {journal} {Annual Review of Fluid
  Mechanics}\ }\textbf {\bibinfo {volume} {36}},\ \bibinfo {pages} {413}
  (\bibinfo {year} {2004})}\BibitemShut {NoStop}%
\bibitem [{\citenamefont {Glaz}\ \emph {et~al.}(2017)\citenamefont {Glaz},
  \citenamefont {Mezi\'{c}}, \citenamefont {Fonoberova},\ and\ \citenamefont
  {Loire}}]{Glaz_2017}%
  \BibitemOpen
  \bibfield  {author} {\bibinfo {author} {\bibfnamefont {B.}~\bibnamefont
  {Glaz}}, \bibinfo {author} {\bibfnamefont {I.}~\bibnamefont {Mezi\'{c}}},
  \bibinfo {author} {\bibfnamefont {M.}~\bibnamefont {Fonoberova}},\ and\
  \bibinfo {author} {\bibfnamefont {S.}~\bibnamefont {Loire}},\ }\bibfield
  {title} {\bibinfo {title} {Quasi-periodic intermittency in oscillating
  cylinder flow},\ }\href
  {https://doi.org/https://doi.org/10.1017/jfm.2017.530} {\bibfield  {journal}
  {\bibinfo  {journal} {Journal of Fluid Mechanics}\ }\textbf {\bibinfo
  {volume} {828}},\ \bibinfo {pages} {680} (\bibinfo {year}
  {2017})}\BibitemShut {NoStop}%
\end{thebibliography}
\end{document}